\newif\ifpaper
\paperfalse

\ifpaper
\documentclass[a4paper,referee]{aa}
\else
\documentclass[a4paper]{aa}
\fi

\usepackage[a4paper]{geometry}
\usepackage{hyperref}
\usepackage{booktabs}
\usepackage{pdflscape}
\usepackage{graphicx}
\usepackage{natbib}
\usepackage{ifthen}
\usepackage{xspace}
\usepackage{titlesec}
\usepackage{adjustbox}

\hypersetup{
pdftitle={LVT151012 and INTEGRAL},
pdfsubject={LVT151012 and INTEGRAL},
pdfauthor={Savchenko et al},
pdfkeywords={LVT151012, INTEGRAL}
}

\newcommand{\ifdiff}[2]{
  \ifthenelse{\equal{#1}{#2}}
    {#1}
    {#1, #2}
}

\newif\ifdebug
\debugfalse

\newif\ifchanges
\changesfalse

\usepackage{color}

\newcommand{\change}[1] {\ifchanges
                        \textbf{#1}
                     \else
                      #1
                     \fi}

\newcommand{\ecs}[1] {{#1}~erg~cm$^{-2}$~s$^{-1}$}
\newcommand{\ec}[1] {{#1}~erg~cm$^{-2}$}

\newcommand{\software}[1] {\textit{#1}}

\newcommand{\lvtevent} {LVT151012\xspace}

\def\la{\mathrel{\mathpalette\fun <}}

\def\fun#1#2{\lower3.6pt\vbox{\baselineskip0pt\lineskip.9pt
\ialign{$\mathsurround=0pt#1\hfil##\hfil$\crcr#2\crcr\sim\crcr}}}

\begin{document}

\title{INTEGRAL IBIS, SPI, and JEM-X observations of \lvtevent}

\authorrunning{V. Savchenko et al.}
\titlerunning{INTEGRAL observations of \lvtevent}
\offprints{V. Savchenko}

\author{V.~Savchenko\inst{\ref{APC},\ref{ISDC}}
	\and  
	A.~Bazzano\inst{\ref{Roma}}
	\and
	E.~Bozzo\inst{\ref{ISDC}}
	\and 
	 S.~Brandt\inst{\ref{DTU}}
	 \and
         J.~Chenevez \inst{\ref{DTU}}
         \and
	  T.~J.-L.~Courvoisier\inst{\ref{ISDC}}
	  \and
		R.~Diehl\inst{\ref{MPE}}
		\and 
		C. Ferrigno\inst{\ref{ISDC}}
		\and 
		L.~Hanlon\inst{\ref{UCD}}
		\and 
		A.~von~Kienlin\inst{\ref{MPE}}
		\and
		E.~Kuulkers\inst{\ref{ESTEC}}
		\and 
		P.~Laurent\inst{\ref{APC},\ref{IRFU}}
		\and 
		F.~Lebrun\inst{\ref{IRFU}}
                \and
                A.~Lutovinov\inst{\ref{SRI_russia},\ref{MIPT_russia}}
                \and
                A.~Martin-Carrillo\inst{\ref{UCD}}
     	        \and
                S.~Mereghetti\inst{\ref{Milano}}
    	        \and
                L.~Natalucci \inst{\ref{Roma}}
		\and
		J.~P.~Roques\inst{\ref{IRAP}}
		\and
		T.~Siegert\inst{\ref{MPE}}
                \and
                R.~Sunyaev\inst{\ref{SRI_russia},\ref{MPA}}
		\and
		P.~Ubertini\inst{\ref{Roma}}
	}

\institute{
APC, AstroParticule et Cosmologie, Universit\'e Paris Diderot, CNRS/IN2P3, CEA/Irfu, Observatoire de Paris
Sorbonne Paris Cit\'e, 10 rue Alice Domont et L\'eonie Duquet, 75205 Paris Cedex 13, France. \label{APC}
\and
ISDC, Department of astronomy, University of Geneva, chemin d'\'Ecogia, 16 CH-1290 Versoix, Switzerland \label{ISDC}
\and
INAF, IASF-Milano, via E.Bassini 15, I-20133 Milano, Italy \label{Milano}
\and
INAF-Institute for Space Astrophysics and Planetology, Via Fosso del Cavaliere 100, 00133-Rome, Italy \label{Roma}
\and
DTU Space - National Space Institute Elektrovej - Building 327 DK-2800 Kongens Lyngby Denmark \label{DTU}
\and
Max-Planck-Institut f\"{u}r Extraterrestrische Physik, Garching, Germany \label{MPE}
\and
Space Science Group, School of Physics, University College Dublin, Belfield, Dublin 4, Ireland \label{UCD}
\and
European Space Research and Technology Centre (ESA/ESTEC), Keplerlaan 1, 2201 AZ Noordwijk, The Netherlands \label{ESTEC}
\and
DSM/Irfu/Service d'Astrophysique, Bat. 709 Orme des Merisiers CEA Saclay, 91191 Gif-sur-Yvette Cedex, France \label{IRFU}
\and
Space Research Institute of Russian Academy of Sciences, Profsoyuznaya 84/32, 117997 Moscow, Russia  \label{SRI_russia}
\and
Moscow Institute of Physics and Technology, Institutskiy per. 9, Dolgoprudny, Moscow Region, 141700, Russia\label{MIPT_russia}
\and
Universit\'e Toulouse; UPS-OMP; CNRS; IRAP; 9 Av. Roche, BP 44346, F-31028 Toulouse, France \label{IRAP}
\and
Max Planck Institute for Astrophysics, Karl-Schwarzschild-Str. 1, Garching b. Munchen D-85741, Germany \label{MPA}
\and
European XFEL GmbH, Albert-Einstein-Ring 19, 22761, Hamburg, Germany \label{XEFL}
}
\date{Received ---; accepted ---}
 
\abstract{
During the first observing run of LIGO, two gravitational wave events
and one lower-significance trigger (\lvtevent) were reported by the
LIGO/Virgo collaboration.  At the time of \lvtevent, the INTErnational
Gamma-Ray Astrophysics Laboratory (INTEGRAL) was pointing at a region
of the sky coincident with the high localization probability area of
the event and thus permitted us to search for its electromagnetic
counterpart (both prompt and afterglow emission).  The imaging
instruments on-board INTEGRAL (IBIS/ISGRI, IBIS/PICsIT, SPI, and the
two JEM-X modules) have been exploited to attempt the detection of any
electromagnetic emission associated with \lvtevent over 3 decades in
energy (from 3~keV to 8~MeV). The omni-directional instruments
on-board the satellite, i.e. the SPI-ACS and the IBIS/Veto,
complemented the capabilities of the IBIS/ISGRI and IBIS/PICsIT for
detections outside their imaging field of view in order to provide an
efficient monitoring of the entire \lvtevent localization region at
energies above 75\,keV.  We did not find any significant transient
source that was spatially and/or temporally coincident with
\lvtevent, obtaining 
tight upper limits on the associated hard X-ray and $\gamma$-ray
radiation.  For typical spectral models, the upper limits on the
fluence of the emission from any 1~s long-lasting counterpart
of \lvtevent ranges from
$F_{\gamma}=$\ec{3.5$\times$10$^{-8}$}
(20~--~200~keV), within the field of view of the imaging
instruments, to
$F_{\gamma}$=\ec{7.1$\times$10$^{-7}$}
(75~--~2000~keV), considering the least favorable
location of the counterpart for a detection by the omni-directional
instruments. These results can be interpreted as a tight constrain on
the ratio of the isotropic equivalent energy released in the
electromagnetic emission to the total energy of the gravitational
waves:
$E_{75-2000~keV}/E_{GW}<$4.4$\times$10$^{-5}$. Finally,
we provide an exhaustive summary of the capabilities of all
instruments on-board INTEGRAL to hunt for $\gamma$-ray counterparts of
gravitational wave events, exploiting both \change{serendipitous} and pointed
follow-up observations. This will serve as a reference for all future
searches.}

\maketitle

\section{Introduction}
\label{sec:introduction}

Gravitational waves (GWs) were predicted as a natural consequence of general
relativity \citep{relativity}, but until recently only indirect
evidence of their existence was found by measuring the time evolution
of orbital parameters in binary pulsars
\citep{gw_indirect,kramer2006}. The first direct detection of
GWs has been achieved with the discovery of GW150914
\citep{triggerpaper} during the first science run of the Advanced
LIGO interferometer (O1). This was followed by the observation of another
high-significance event, GW151226 \citep{GW151226}. 

A third possible GW event, \lvtevent, was detected during O1 with a
low statistical significance of 1.7~$\sigma$ \change{or about 4.5\% in the
O1} \citep[corresponding to a false alarm rate of once in 2.7
years;][]{ligoO1}.  The trigger occurred on UTC 2015-10-12~09:54:43,
and had signal-to-noise ratio (SNR) of 9.7 (to be compared to 13 of
GW151226 and 23.7 of GW150914). Considering the relatively low
detection significance, this event has not been confidently classified
by the LIGO/Virgo collaboration as a GW burst, but rather as
a \textit{LIGO/Virgo Transient}, possibly associated with an
astrophysical event. In this case, it would correspond to a merger of
two black holes of 23$^{+18}_{-6}$~M$_{\odot}$ and
13$^{+4}_{-5}$~M$_{\odot}$ at a distance of 1000$^{+500}_{-500}$~Mpc
\citep{ligoO1}.  \lvtevent has been localized within an area of about 
1600~deg$^2$ (90\% confidence level), consisting of two elongated regions 
spanning over 60 degrees each. Owing to their characteristic shapes, we   
refer to them in the following as the southern and northern arc LIGO regions.  

Electromagnetic counterparts of GW events are mainly expected if at least  
one neutron star is involved \citep[see e.g.][]{voss2003,Siellez2013,Patricelli2016}.  However, it cannot be excluded
that the merging of black holes could
produce an electromagnetic signal under particular conditions (see Sect.~\ref{sec:discussion}).  
To promote the searches for possible electromagnetic counterparts of
GW events, the LIGO/Virgo collaboration distributes
near real time alerts to selected teams who have signed a memorandum
of understanding.  These alerts contain a localization
probability map for each event and have led so far to a massive follow-up 
campaign of GW150914 and
GW151226 \citep{triggerempaper,triggerempaper2,GW151226_GBM,GW151226_JGEM,GW151226_neutrino,
GW151226_neutrino2,GW161226_Calet,GW161226_Decam,GW161226_PAO}.
In the case of \lvtevent,\ the localization was distributed to the
community only 6 months after the event. No extensive follow-up observations 
have thus been performed and only serendipitous data-sets are available from all relevant 
facilities. 

The INTErnational Gamma-Ray Astrophysics
Laboratory \citep[INTEGRAL;][]{integral} team participated to the search for electromagnetic
counterparts of GW150914 and provided the tightest upper limits for the whole
LIGO localization region above 75\,keV, challenging the reported evidence for a 
temporarily coincident $\gamma$-ray event originated from the merging of two black holes 
\citep{savchenko16,gbmpaper,Greiner2016}.  
At the time of GW151226, INTEGRAL was not taking scientific data due to
proximity of one of the perigee passages\footnote{Note that the INTEGRAL satellite 
orbital period was reduced from 3 to 2.6 sidereal days in January 2015.}. 
Slightly before and after each of these passages, 
all instruments on-board INTEGRAL are switched off to prevent damages while crossing the 
Earth radiation belts, thus limiting the observational efficiency of the observatory (``duty cycle'') 
to 85\%. 

We report in this paper the results of our serendipitous search for an
electromagnetic counterpart of the third possible GW event, \lvtevent. 
\change{In Sect.~\ref{sec:observations}, we first introduce
the INTEGRAL general capabilities and then 
describe the observations performed at the time of \lvtevent. 
Then, we report the upper limits obtained with each instrument 
on any possible hard X-ray/$\gamma$-ray counterpart of \lvtevent.
We detail the characteristics
of each instrument relevant for our purpose in the Appendix. }
This section provides an extensive
summary of the capabilities of all INTEGRAL instruments to perform similar searches, 
serving as a references for future studies of transient events. 
In Sect.~\ref{sec:results}, we
combine together the different results to achieve the most stringent upper
limit on the electromagnetic emission from \lvtevent.  
We discuss our results in the context of different models for the 
electromagnetic counterparts of GW events, illustrating the potential of our search for 
the detection of prompt and afterglow-like emission from similar sources (Sect.~\ref{sec:discussion}). 
We draw our conclusions in Sect.~\ref{sec:conslusions}.

\section{Observations}
\label{sec:observations}

INTEGRAL was designed to perform observations in the hard~X-ray~/~soft
$\gamma$-ray energy range over a $\sim$30$\times$30~deg field of view (FoV) with
two main instruments: IBIS, optimized for high angular
resolution \citep{ibis}, and SPI providing high spectral resolution
\citep{spi}. In addition, the JEM-X and
OMC instruments provide simultaneous monitoring in the central part of
the FoV in the soft X-ray and optical domains, respectively
\citep{jemx,omc}.
IBIS and SPI are equipped with active shields to reduce the cosmic-ray
induced background.  The event count-rate provided by these shields
can be used to detect transient $\gamma$-ray phenomena over the
full sky, making the Anti Coincidence Shield (ACS) of SPI and the
Veto system of IBIS two $\gamma$-ray detectors with a competitive sensitivity for such purpose. 

The observational mode of INTEGRAL foresees several pointings lasting roughly 1~hour and 
with an aim position that is changed by a few degrees from one pointing to the other (``dithering strategy'') in order 
minimize the systematics of the coded mask imaging. The imaging instruments allow us to reach deeper sensitivities 
(even though in a much more limited region of the sky) compared 
to the omni-directional instruments, which can monitor the entire sky but are 
typically affected by a higher background featuring much less controllable variability. 
As we detail later, some of the imaging instruments are also able to provide detections for events 
occurring outside their FoVs at the expenses of imaging capabilities. 

\lvtevent took place approximately 5 minutes before the end of the pointing 159700850010, the last scientific one in
 the INTEGRAL revolution 1597.  This pointing started on 2015 October
12 at 2015-10-12~08:59:14 (UTC) and was concluded on the same day at
09:59:20 (UTC). At 09:58:48 (UTC), SPI was configured to enter the
radiation belts and thus the SPI-ACS and SPI data acquisition was
interrupted at 09:59:20 (UTC). \change{In the current configuration,
SPI enters stand-by mode slightly before the radiation belt passage.}
IBIS and JEM-X collected up to 230 seconds of additional science data
during the following pointing 159700860010, but the observing
conditions were not optimal due to the enhanced background
variability. We thus comment on these data separately.  The aim point
of the imaging instruments during the pointing 159700850010 was at
RA~=~250.7, Dec~=~--45.9, i.e. close to the most probable localization
of \lvtevent within the southern arc LIGO region (RA~=~245.2,
Dec~=~--32.9). At least for a fraction of this region (up to 20\%,
depending on the instrument), we could exploit for the first time the
high sensitivity of the observations obtained within the FoVs of the
imaging instruments (JEM-X, IBIS/ISGRI, IBIS/PICsIT, and SPI) to
search for both an impulsive event and an afterglow-like emission. The
FoV of JEM-X, IBIS/ISGRI, IBIS/PICsIT, and SPI cover an area
of 130~deg$^2$, 900~deg$^2$,
900~deg$^2$, and 1225~deg$^2$, respectively\footnote{These FoVs coverages are
provided at 0\% coding efficiency.}. The remaining part of the southern arc and the northern arc of the LIGO localization region 
were only covered with lower sensitivity by the omni-directional instruments and through the out-of-FoV observations with the 
IBIS/ISGRI and IBIS/PICsIT. We verified that at the time of the GW trigger  
the region of the sky occulted by the Earth was nearly 5~deg in radius for INTEGRAL but it did not overlap with 
any portion of the LIGO localization region of \lvtevent where the detection probability of the GW event was significantly higher than zero. 

Given the different availabilities of the INTEGRAL instruments specified above, we performed two independent types of searches 
in order to identify a possible $\gamma$-ray counterpart to \lvtevent: 

\begin{enumerate}
\item search for the prompt burst emission of \lvtevent (``impulsive event''). In this case, we exploited both the data collected by the 
imaging (within and outside their FoV) and omni-directional instruments in a time interval 
spanning from 30 seconds before to 30 seconds after the time of the LIGO trigger. Two different 
possibilities were considered for the spectral energy distribution of the event: (i) the \emph{hard} Comptonized
  model that describes the typical short GRB spectrum, parametrized by a cutoff powerlaw with 
  E$_{peak}$~=~600~keV and
  $\alpha$~=~-0.5 \citep[these values are close to the average of the measurements obtained from 
  all GBM short GRBs;][]{gruber14}; (ii) the Band model usually adopted to describe \emph{long-soft}
  GRBs with representative parameters E$_{peak}$ = 300~keV,
  $\alpha$~=~-1, and $\beta$
  =-2.5. We assumed the durations of 1~s and 8~s in order to be representative of the cases of 
  short and long GRBs, respectively. The specific choice of 8~s binning is driven by the readout time of the IBIS/Veto
  light curve (see Sect.~\ref{sec:ibisveto}).

\item search for the long-lasting afterglow of \lvtevent. In this case, we concentrated on the detection 
of a hard X-ray/$\gamma$-ray emission in the FoV of  imaging instruments that could resemble a GRB afterglow. 
We  assumed a power-law decay of the emission intensity over time and a relatively soft spectral energy distribution 
($F_E \sim E^{-2}$).
\end{enumerate}

In the following sub-sections, we describe in detail the performances of
each INTEGRAL instrument for the above searches and derive separated upper limits on the flux of any 
possible electromagnetic counterpart of \lvtevent. 

\subsection{JEM-X results} \label{sec:jemx_results}

Both JEM-X modules \change{(see also Appendix~\ref{sec:jemx})} were nominally
operated during the entire INTEGRAL pointing 159700850010, securing a
reliable gain determination. We also verified that no particularly
large background variations connected to the solar activity affected
the JEM-X data collected in the time interval spanning from 300~s
before the \lvtevent detection by LIGO and the end of the INTEGRAL
pointing.

\ifpaper
\newgeometry{left=1cm}
\begin{landscape}
\fi

\begin{figure*}
  \centering 
    \makebox[\linewidth]{
\ifpaper
  \includegraphics[width=1.1\columnwidth, angle = 0]{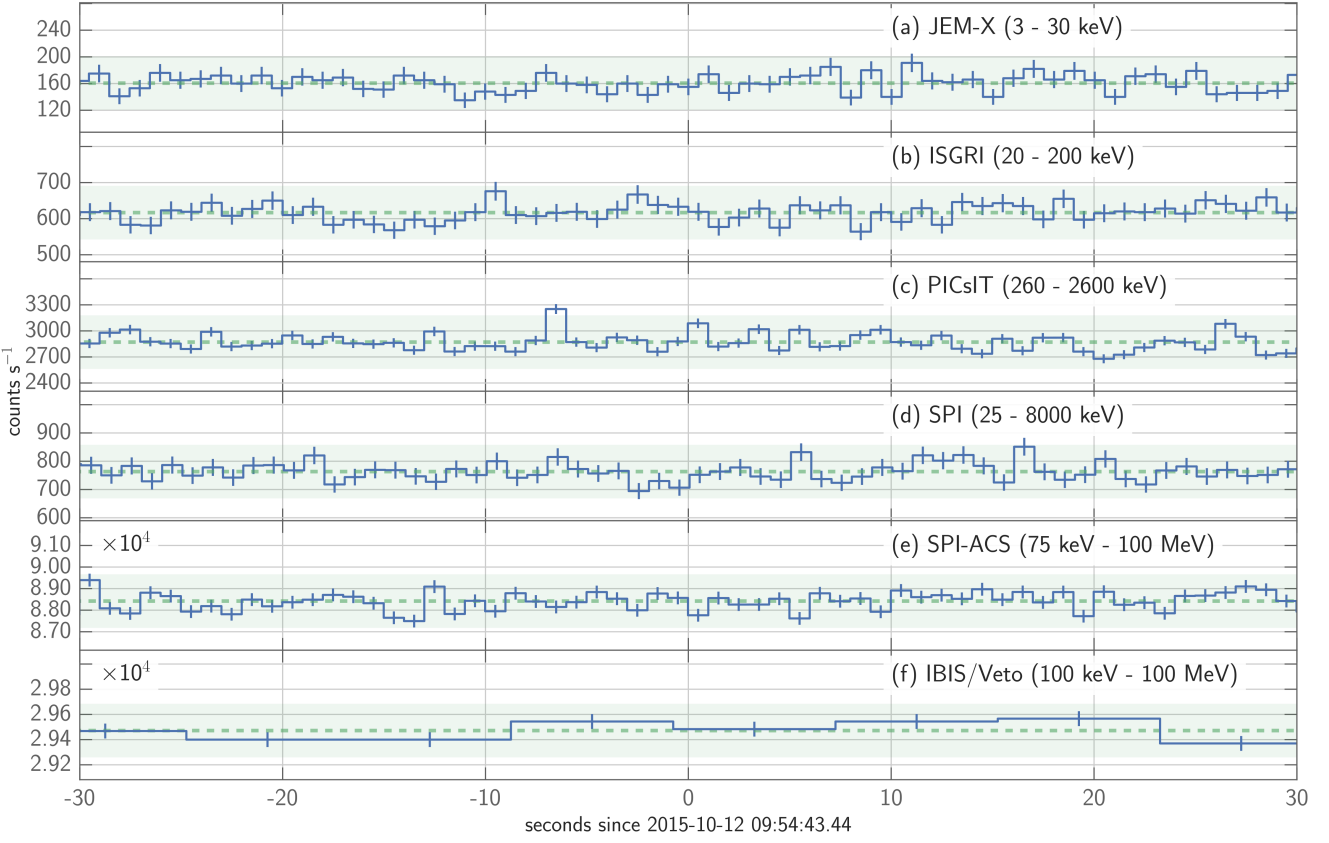}
\else
  \includegraphics[width=0.95\textwidth, angle = 0]{lvt151012_multilc_patched.eps}
\fi
    }
  \caption{Light curves obtained from the total detector count-rates of different instruments on-board INTEGRAL and zoomed 
  around the time of  \lvtevent\ trigger provided by the LIGO/Virgo collaboration (see Sect.~\ref{sec:introduction}). 
  All light curves are binned with a time resolution   of 1~s, the only exception being the one obtained from 
the IBIS/Veto which has an intrinsic time resolution of 8~s 
  (see Sect.~\ref{sec:ibisveto}). We also indicated in the figure the count-rates of the different instrument backgrounds 
  (dashed lines) and the expected level of random fluctuations at 3~$\sigma$ confidence level (shaded regions).}
  \label{fig:total_rate_lc} 
\end{figure*}

\ifpaper
\end{landscape}
\fi

\begin{figure}
  \centering 
  \resizebox{\hsize}{!}{\includegraphics[trim={0cm 0cm 0cm 0cm},width=1.\columnwidth, angle =0]{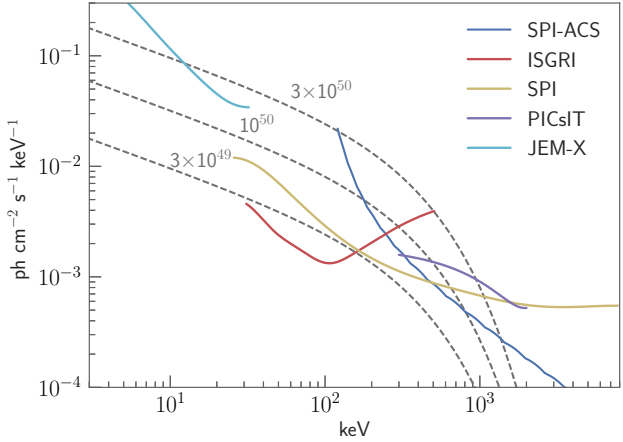}}
  \caption{Plot of the broad-band continuum sensitivity ($\Delta$E=E/2) 
  of the different instruments on-board INTEGRAL as a function of  energy for an on-axis  
  detection of a 1~s-long impulsive event at 3~$\sigma$ confidence level during the time interval spanning from  
  T$_0$-30~s to T$_0$+30~s. Dashed curves correspond to the case of an event at 1~Gpc featuring a spectral energy 
  distribution typical of a hard GRB (Comptonized model with parameters $\alpha=-0.5$, 
  $E_{peak}=$600$~$keV) 
    and releasing a total of 3$\times$10$^{49}$~ergs, 10$^{50}$~ergs, and 3$\times$10$^{50}$~ergs in the 75-2000~keV energy band.}
  \label{fig:spec_sens_fov_short}
\end{figure}

The inspection of the count-rate collected by JEM-X during the 30 seconds 
preceding and following the onset of \lvtevent did not show any convincing presence of 
 impulsive events (see Figure~\ref{fig:total_rate_lc}a). To derive an accurate upper limit 
on the X-ray flux of the non-detected counterpart to the GW source, we evaluated the instrument sensitivity by 
using 
an observation of the Crab pulsar and nebula carried out earlier in the same revolution (1597). 
Combining the detector count-rates from both JEM-X modules, we found that the limiting detectable 3-35~keV 
fluence of a hard 1~s-long (soft 8~s-long) event in a location compatible with that of \lvtevent and within the instrument FoV 
is \ecs{4.5$\times$10$^{-8}$} 
(\ecs{1.6$\times$10$^{-7}$}) at 3$\sigma$ confidence level 
(see also Figure~\ref{fig:spec_sens_fov_short} for the energy dependency of the instrument sensitivity). 

\begin{figure}
  \centering 
  \resizebox{\hsize}{!}{\includegraphics[trim={0cm 0cm 0cm 0cm},width=\columnwidth, angle = 0]{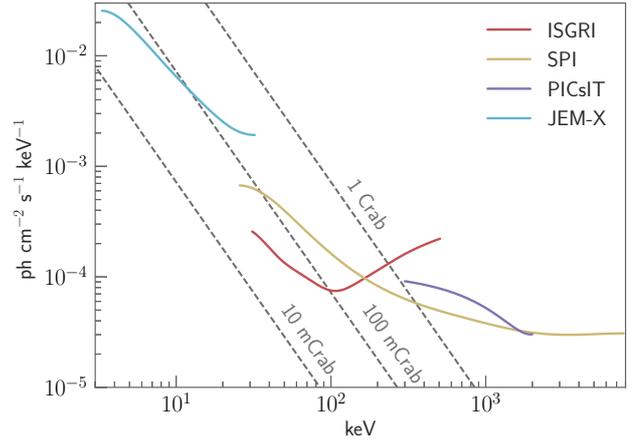}}
  \caption{Plot of the broad-band continuum sensitivity ($\Delta$E=E/2) 
  of the different instruments on-board INTEGRAL as a function of  energy for a 
  detection of an on-axis source at 3~$\sigma$ confidence level during the time interval spanning from a minimum of 
  T$_0$-30~s to a maximum of T$_0$+287~s (depending on instrument). 
  The dashed lines correspond to the case in which the source to be detected is characterized by a power-law 
  shaped spectrum with a photon index of $-2$ and an X-ray flux below 100~keV equal to 1~Crab, 100~mCrab, and 10~mCrab, 
  respectively.}
  \label{fig:spec_sens_fov_long}
\end{figure}

\begin{figure}
  \centering
  \resizebox{\hsize}{!}{
  \includegraphics[width=0.9\columnwidth]{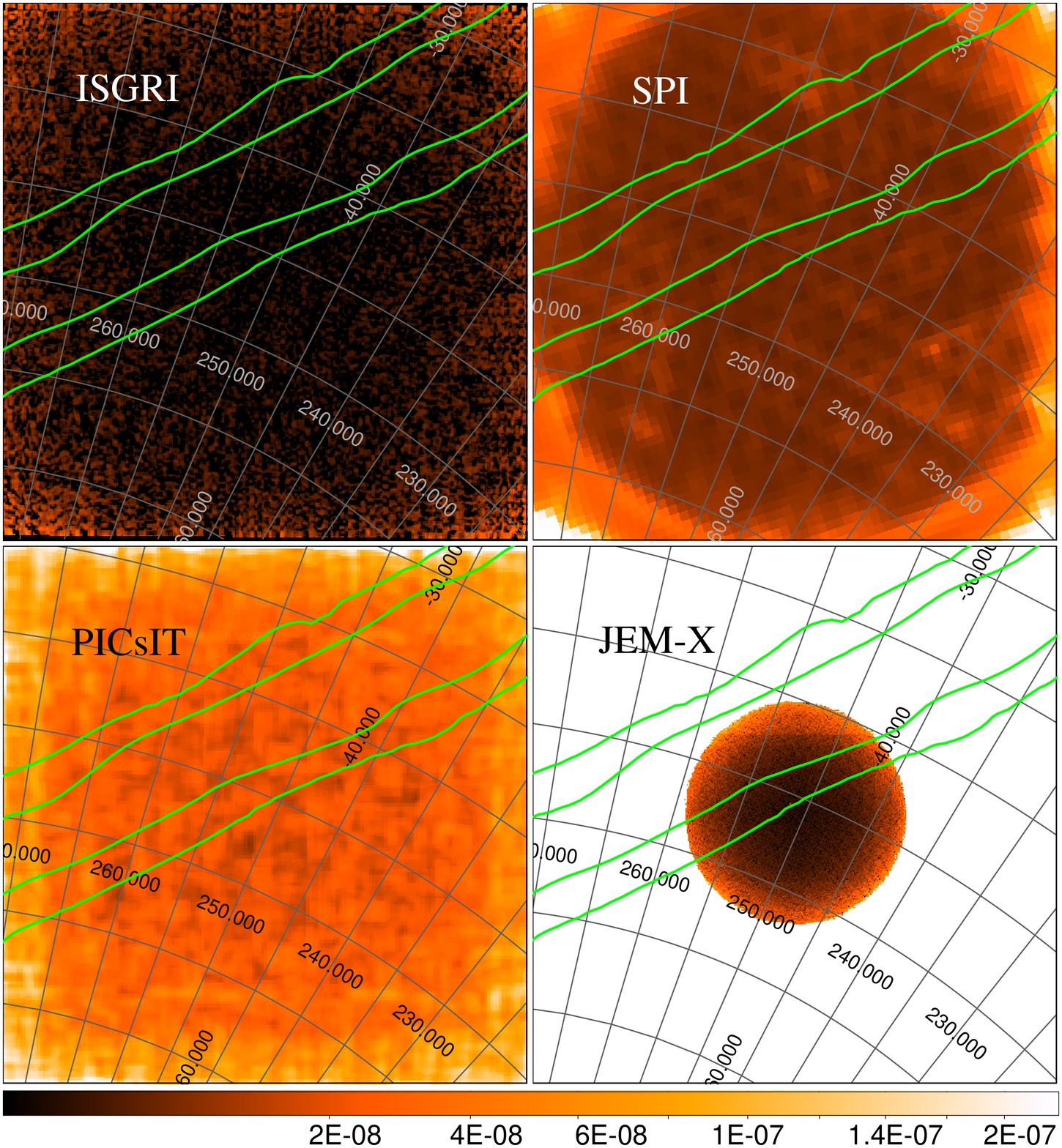}}
  \caption{Sensitivity maps within the INTEGRAL IBIS/ISGRI (20~--~200 keV), IBIS/PICsIT (260~~--~2000~keV), SPI (25~--~8000~keV), 
  and the combined JEM-X1 and JEM-X2 modules (3.~--~35.~keV) FoVs during the time interval spanning from a minimum of T$_{0}$-30~s to 
    a  maximum of T$_{0}$+287~s (depending on instrument). The units reported below the color bar are \ecs{}. 
    We also show in green the 90\% and 50\% confidence level contours of the LIGO \lvtevent localization region.}
  \label{fig:sens_4fov}
\end{figure}

To search for a possible GRB afterglow-like event, we built the two JEM-X 
images in the time interval spanning from 30 seconds before the LIGO detection up to 287~s after. 
We made use of the standard OSA software (v10.2) distributed by the ISDC \citep{courvoisier03} and combined the 
two images to achieve the highest possible sensitivity.  We did not find any convincing source 
in the summed image and, assuming a powerlaw shaped energy spectrum with $\alpha$=-2, we 
estimated the most stringent 3$\sigma$ upper limit on the 3-35~keV X-ray flux of a source observed on-axis at 
\ecs{1.4$\times$10$^{-9}$}  
(see also Figs.~\ref{fig:spec_sens_fov_long} and \ref{fig:sens_4fov} for the upper limit dependency 
on source position within the JEM-X FoV).  

As anticipated in Sect.~\ref{sec:observations}, the two JEM-X modules
collected about 230~s of additional data in the pointing 159700860010.
An inspection of these data revealed that they were strongly affected by
variable background. Thus, we do not discuss these data any further.

\subsection{IBIS/ISGRI results}\label{sec:isgri_results}

We searched for impulsive events in the full detector light curves (neglecting the 
imaging capabilities, \change{see Appendix~\ref{sec:isgri}}) covering the 20--200~keV energy range and the time interval 
spanning from T$_0$-30~s to T$_0$+30~s (where T$_0$ is the time of the LIGO trigger; see Figure~\ref{fig:total_rate_lc}b).
No significant detection was found and we estimated the most stringent 3$\sigma$ 
upper limit on the 20~--~200~keV flux of a 1~s-long (8~s-long) event of  
\ecs{3.5$\times$10$^{-8}$} (\ecs{7.5$\times$10$^{-8}$}), assuming the localization of the 
event in the center of the instrument FoV and a typical hard (soft) GRB-like spectrum (see Table~\ref{tab:prompt}). 
The energy dependence of the upper limit for the 1~s-long hard event is shown in Figure~\ref{fig:spec_sens_fov_short}. 
ISGRI response and sensitivity coverage for events occurring outside its FoV are illustrated in 
Sect.~\ref{sec:shortevent} and Figure~\ref{fig:sky_sens_4} (see also Figure~\ref{fig:isens_zenith_hard}).   

We searched for GRB afterglow-like events in the ISGRI image 
extracted with the OSA 10.2 software in the time interval 
ranging from T$_0$-30~s to T$_0$+287~s.  
We did not find convincing evidence of any new source that could be associated with the 
GW event within the ISGRI FoV. The most stringent upper limit on the corresponding flux 
is derived within the instrument fully coded FoV and corresponds to 
\ecs{7.4$\times$10$^{-10}$} 
(assuming a powerlaw spectrum with a slope of $-2$). 
The dependence of  ISGRI sensitivity on  energy during the time interval 
T$_0$-30~s~--~T$_0$+287~s is shown in Figure~\ref{fig:spec_sens_fov_long}.

About 230~s of additional ISGRI data were accumulated starting on 2015
October 12 at during pointing 159700860010.  During this time
interval, the instrument was affected by larger background and we
could not find any evidence for a convincing new source detection in
the corresponding image. We also checked that combining these
observations with the previously used exposure resulted only in a
marginally improved sensitivity. Thus, we  neglected ISGRI data in
pointing 159700860010 while combining the results from all INTEGRAL
instruments in Sect.~\ref{sec:results}.

\subsection{IBIS/PICsIT results}

We first built the PICsIT full detector light curves in the energy
range 260-2600~keV with a time resolution of 1~s (see
Figure~\ref{fig:total_rate_lc}c) and estimated the instrument
background (together with its variance) to search for possible
impulsive events within the time interval spanning from T$_0$-30~s to
T$_0$+30~s.  A marginally significant excess (signal-to-noise ratio,
SNR, of 3.9)
was found at T0~-~6.5~s. The
corresponding post-trial significance of this excess
is 2.0,
too low to be considered a realistic detection. We thus set the
3~$\sigma$ upper limit on the fluence of any 1~s-long (8~s-long)
impulsive event not detected by PICsIT within the fully-coded
instrument FoV
at \ec{9.7$\times$10$^{-7}$}
(\ec{1.9$\times$10$^{-6}$}). The
obtained PICsIT sensitivity as a function of energy is shown in
Figure~\ref{fig:spec_sens_fov_short}.  The PICsIT sensitivity to
events occurring outside its FoV is illustrated in
Sect.~\ref{sec:shortevent} (see also
Figure~\ref{fig:isens_zenith_hard}).

We searched for the presence of possible long-lasting GRB
afterglow-like emissions in the PICsIT images (260-2600~keV) that
could be associated with \lvtevent.  No convincing detection was
found. The corresponding results are presented in
Figure~\ref{fig:spec_sens_fov_long} and \ref{fig:sens_4fov}.

\subsection{IBIS Compton mode results} \label{sec:comptonmoderes}

From the measurements of the IBIS Compton-mode background in the vicinity of \lvtevent,
we verified that the Compton kinematics could be
used to reduce by 50\% the background of an on-axis source and about
75\% that of a source located 40~deg off-axis observed in this mode. A limited amount of detected photons 
have a localization accuracy as good as 10\% of the sky. However,  the use of only these photons 
to search for possible counterparts to \lvtevent would dramatically reduce the statistics of any signal that can be 
recorded from any point-like source. The highest sensitivity to the emission from a point source with an unknown position 
based on the Compton kinematics was obtained without applying any event selection, i.e., by using the total IBIS Compton-mode 
count-rate.

As the effective area of IBIS Compton mode is smaller than that of
both ISGRI and PICsIT, the lower number of recorded photons in this
mode cannot provide by definition a more stringent upper limit on the
non-detected counterpart to \lvtevent when compared to what was
discussed before.

\subsection{IBIS/Veto Results} \label{sec:ibisveto_results}

We show in Figure~\ref{fig:total_rate_lc}f the IBIS/Veto light curve close to the 
time of the \lvtevent detection and with a bin time of 8~s. 
No obvious impulsive events are found within a time interval of $\pm$~30~s around the occurrence of
\lvtevent. Similarly to the SPI-ACS, the IBIS/Veto is unable to localize the events and provide any 
energy resolution. Assuming the two representative GRB spectra mentioned in Sect.~\ref{sec:observations} 
and considering all accessible regions of interest for the IBIS/Veto, we determined 3~$\sigma$ 
upper limits on the fluence of any impulsive events in a 8~s time bin 
ranging from \ec{4.4$\times$10$^{-7}$} to
\ec{9.5$\times$10$^{-6}$} (see also
Table~\ref{tab:prompt}).

\subsection{SPI results} \label{sec:spi_results}

We first extracted the total instrument light curve in the 25--8000~keV 
energy range with a time resolution of 1~s in the time interval 
spanning from T$_0$-30~s to T$_0$+30~s. This light curve is shown in 
Figure~\ref{fig:total_rate_lc}d. No clear signatures for the 
presence of significant impulsive events are observed. 

In order to estimate the instrument sensitivity to the detection 
of transient sources through the coded mask imaging within the fully-coded FoV, we first 
performed a maximum likelihood fit in each energy bin by assuming a combined 
contribution to the data from detector background and a single on-axis point source. 
We derived a 3~$\sigma$ upper limit on the fluence of any non-detected 1~s-long event 
during the investigated time interval of 5.7$\times$10$^{-7}$~erg~cm$^{-2}$  
in the 25--8000~keV energy range. We show in Figure~\ref{fig:spec_sens_fov_short} 
the sensitivity of SPI for similar events as a function of energy. 
The instrument sensitivity gradually decreases in the 
partially coded FoV (i.e., for off-axis angles larger than 8
deg), becoming null at 16~deg.

SPI images can also be used to search for GRB afterglow-like events. 
We could not detect any significant new source in the available 
image integrated from T$_0$--30\,s to the end of  the SPI observation in INTEGRAL pointing 159700850010 
(T$_0$+245\,s). We estimated a correspondingly 3~$\sigma$ upper limit 
on the 25--1000\,keV X-ray flux of $2.6\times10^{-9}$~erg~cm$^{-2}$~s$^{-1}$, 
assuming a power-law shaped source spectral energy distribution with a 
photon index of $-2$ and a source located within the instrument fully-coded FoV. 
The dependence of  SPI sensitivity on energy for the detection 
of GRB afterglow-like events is shown in Figure~\ref{fig:spec_sens_fov_long}, while 
the dependence on source location within the instrument FoV is shown 
in Figure~\ref{fig:sens_4fov}.

\subsection{SPI-ACS results}

We extracted the full SPI-ACS detector light curve around the time of  \lvtevent detection 
(from T$_0$-30\,s to T$_0$+30\,s). The light curve is shown in Figure~\ref{fig:total_rate_lc}e and 
does not display any signature of significant impulsive events above the 
instrumental background. 
For the fluence upper limit evaluation, we 
inspected the instrument background variability and calculated the corresponding variance 
in the time interval spanning from T$_0$-300~s to T$_0$+287~s  
(\lvtevent occurred close to the end of  INTEGRAL orbit 
where the instruments are usually hit by the stream of  Earth magnetospheric particles). 
We verified that the background variability was mainly limited to  
a low-frequency noise ($<$~0.1~Hz)  in the time interval of interest. 
As the SPI-ACS has no energy resolution and localization
capabilities, its sensitivity strongly depends on the assumptions on the
source spectral energy distribution and position in the sky. Considering the two representative GRB
spectra mentioned in Sect.~\ref{sec:observations} and scanning the full sky area accessible to the 
SPI-ACS, we determined 3~$\sigma$ upper limits on the fluence of the non-detected 1~s-long (8~s-long) event ranging
from \ec{1.3$\times$10$^{-7}$} (\ec{3.3$\times$10$^{-7}$})
to \ec{7.1$\times$10$^{-7}$} 
(\ec{2.2$\times$10$^{-6}$}; see also Table~\ref{tab:prompt}).

Evaluating SPI-ACS sensitivity as a function of the energy is
particularly challenging, as it depends strongly on the 
spectral models considered. Assuming a number of spectra described through the 
GRB Band model with $\alpha$ ranging from 0 to -2, E$_{peak}$ from 50~keV to
5~MeV, and $\beta$ from -2.5 to -5.5, we computed the function describing the 
maximum photon flux at different energies that each spectrum should have in order \emph{not} to be  
detected by the SPI-ACS in 1~s (at 3~sigma confidence level). 
We use this function to plot a conservative estimate of  SPI-ACS sensitivity in 
Figure~\ref{fig:spec_sens_fov_short}.

\ifpaper
\begin{landscape}
\fi

\begin{figure*}
  \centering 
  \vspace{0.5cm}

\begin{tabular}{cc}

\ifpaper
  \includegraphics[trim={1cm 0cm 0cm 1cm},width=0.4\columnwidth, angle
    =0]{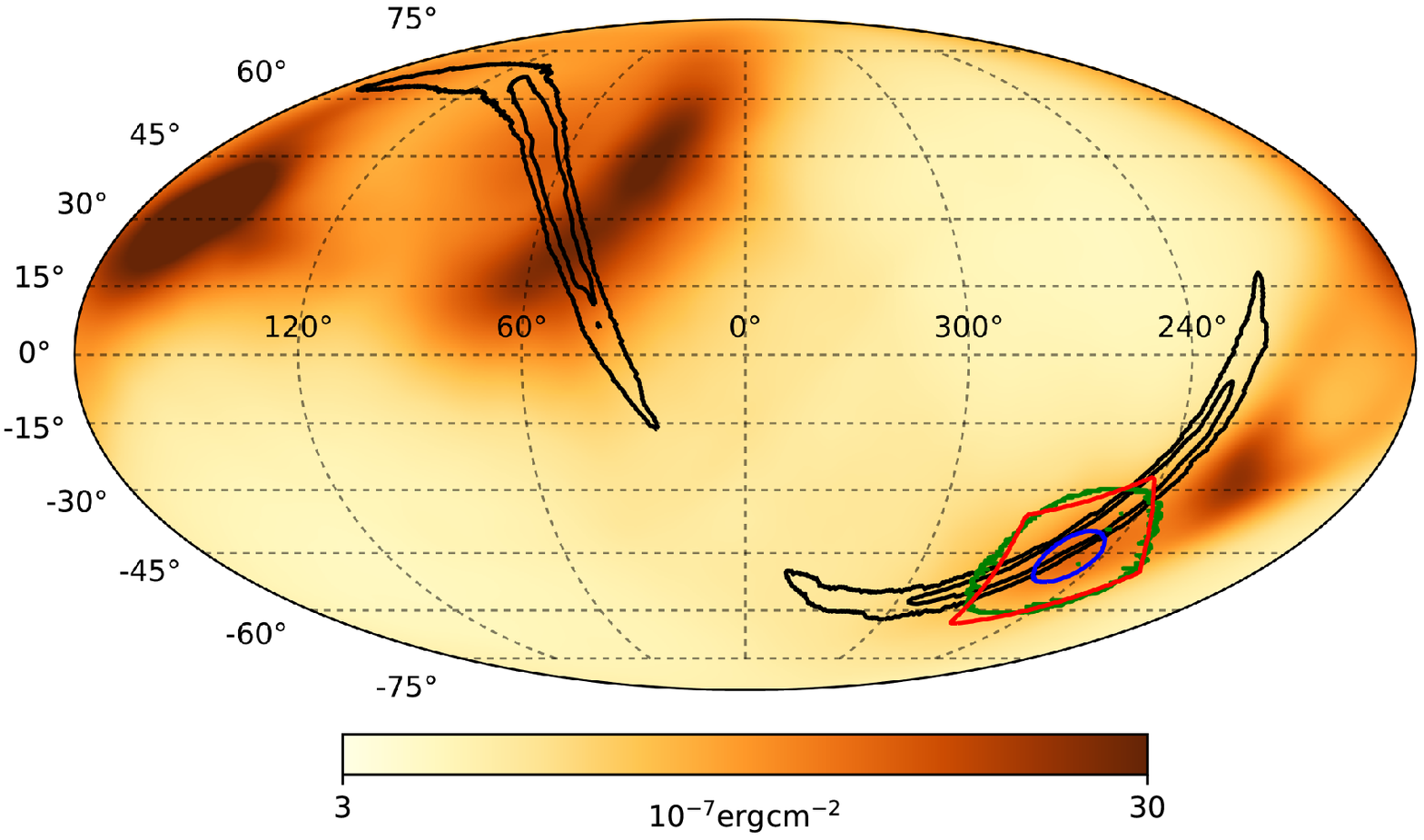} &
  \includegraphics[trim={1cm 0cm 0cm 1cm},width=0.4\columnwidth, angle
    =0]{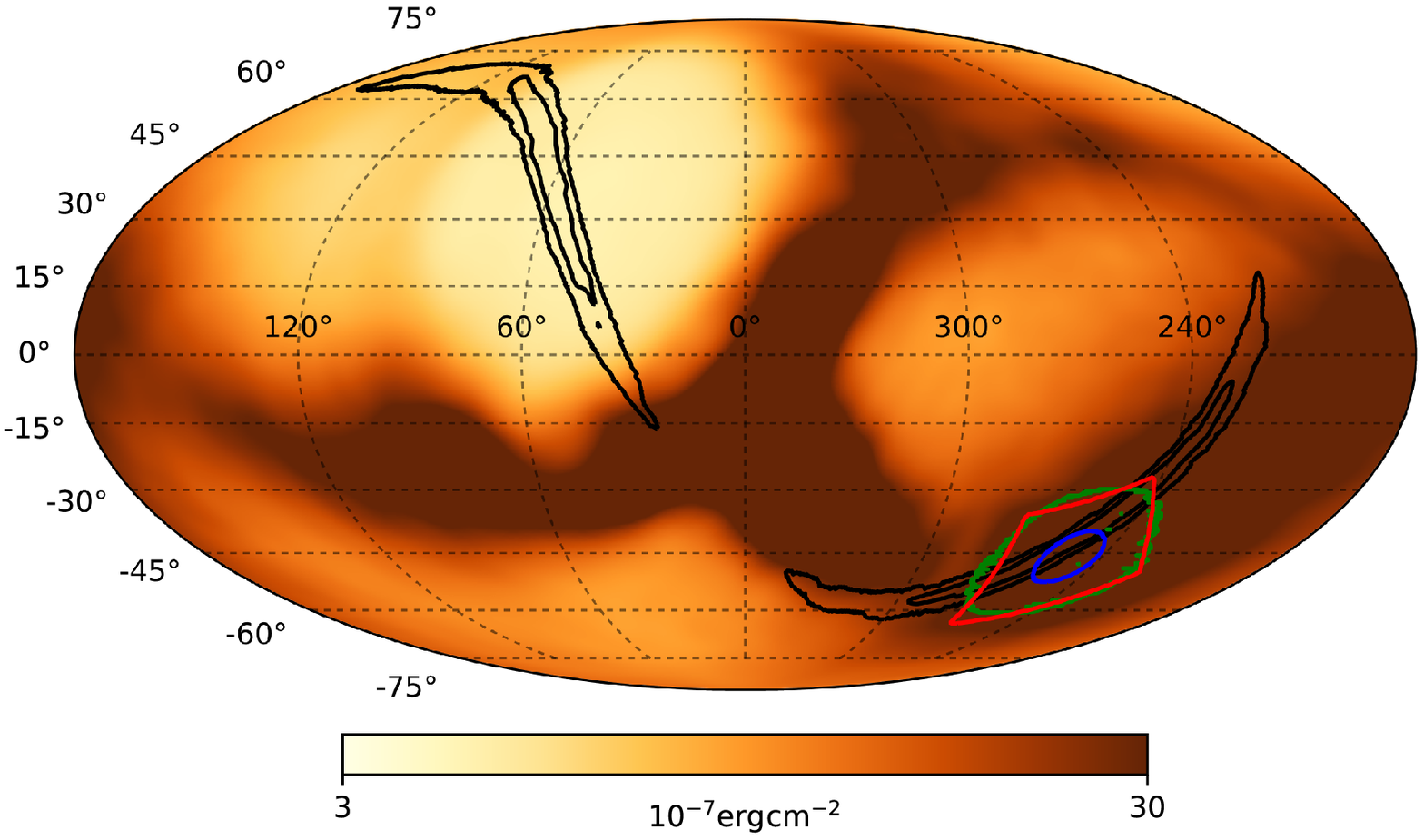} \\
  \includegraphics[trim={1cm 0cm 0cm 1cm},width=0.4\columnwidth, angle
    =0]{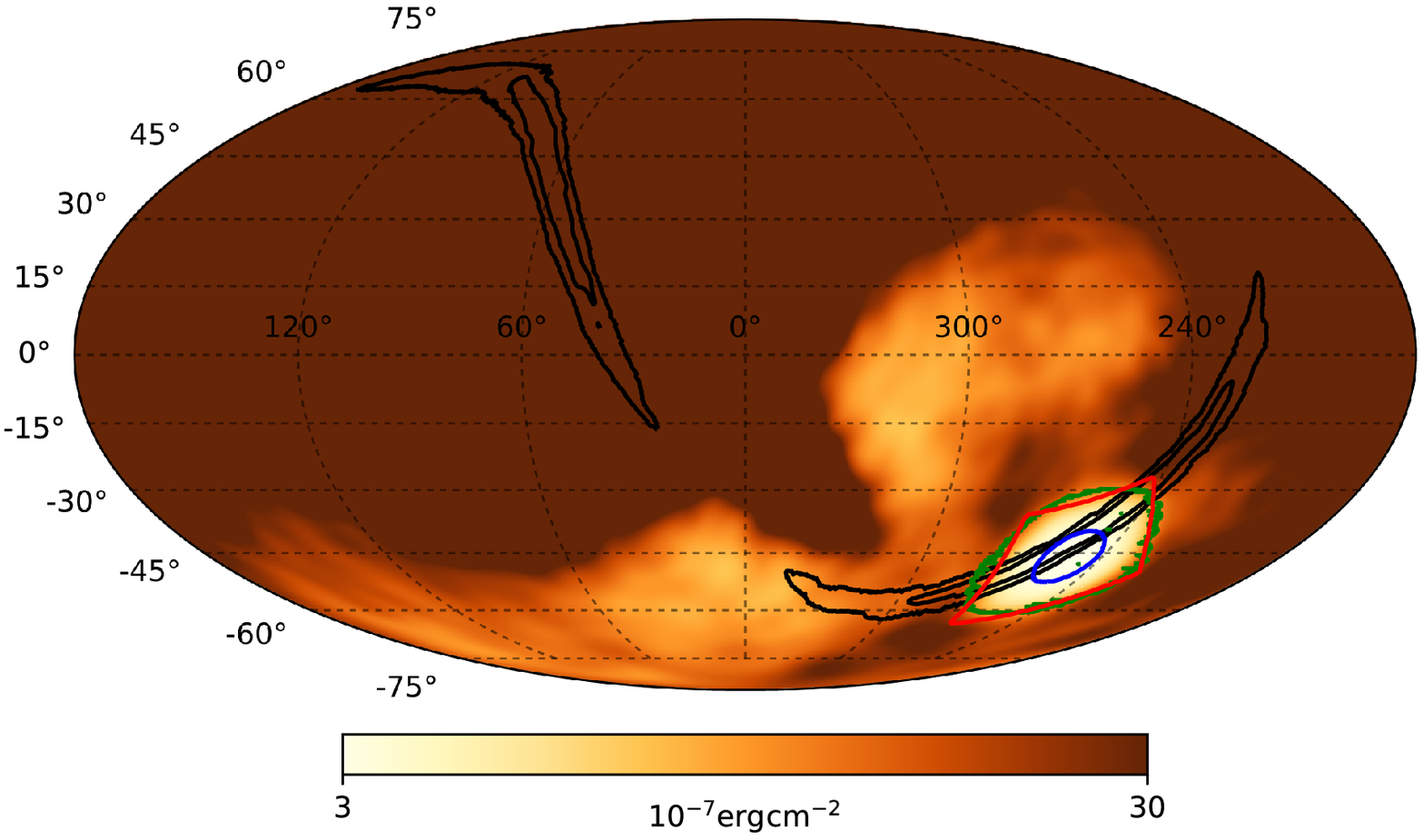} &
  \includegraphics[trim={1cm 0cm 0cm 1cm},width=0.4\columnwidth, angle
    =0]{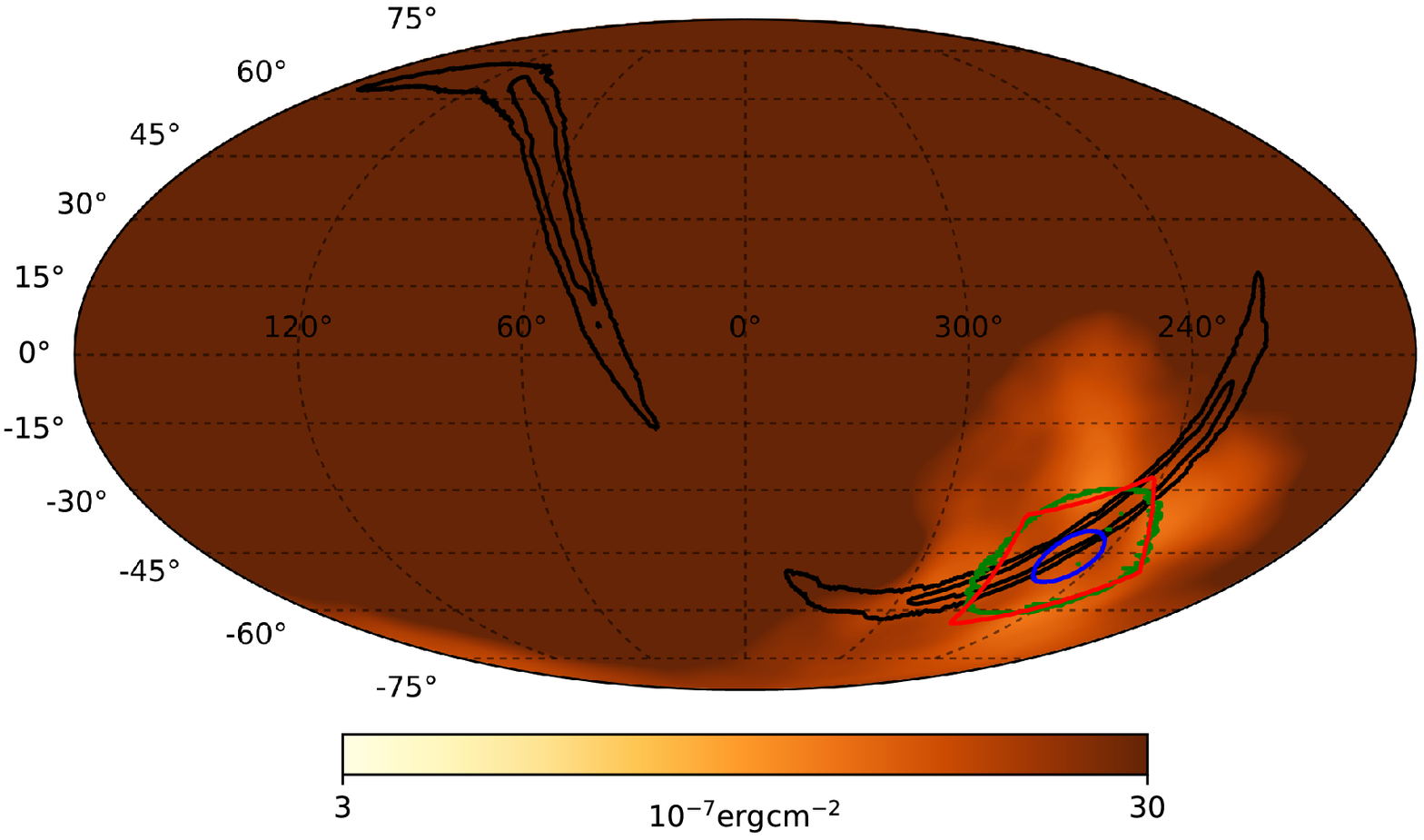} \\
\else
  \includegraphics[trim={1cm 0cm 0cm 1cm},width=\columnwidth, angle
    =0]{sky_sens_acs__8s.eps} &
  \includegraphics[trim={1cm 0cm 0cm 1cm},width=\columnwidth, angle
    =0]{sky_sens_veto__8s.eps} \\
  \includegraphics[trim={1cm 0cm 0cm 1cm},width=\columnwidth, angle
    =0]{sky_sens_isgri__8s.eps} &
  \includegraphics[trim={1cm 0cm 0cm 1cm},width=\columnwidth, angle
    =0]{sky_sens_picsit__8s.eps} \\
\fi

\end{tabular}

  \caption{Sensitivity maps of  SPI-ACS, IBIS/Veto,
    IBIS/PICsIT, and IBIS/ISGRI for the detection of a 8~s-long soft GRB event at 3~$\sigma$ confidence level whose spectral 
    energy distribution is described by a Comptonized model with parameters $\alpha=$-1 and 
    $E_{peak}=$300$~$keV). 
    We represent with black contours the most accurate localization of the GW event 
    (at 50\% and 90\% confidence levels) obtained by the LALInference \citep{ligoO1}. 
    Red, green, and blue contours indicate the sky regions within which the sensitivity of 
    ISGRI, SPI, and JEM-X is not degrading more than a factor of 20 compared to the on-axis value, respectively.}
  \label{fig:sky_sens_4}
\end{figure*}
\ifpaper
\end{landscape}
\fi

\ifpaper
\newgeometry{left=1cm}
\begin{landscape}
\fi

\begin{figure}
  \centering
\ifpaper
  \resizebox{\hsize}{!}{\includegraphics[trim={1cm 1cm 1cm 1cm},width=0.8\columnwidth, angle=0]{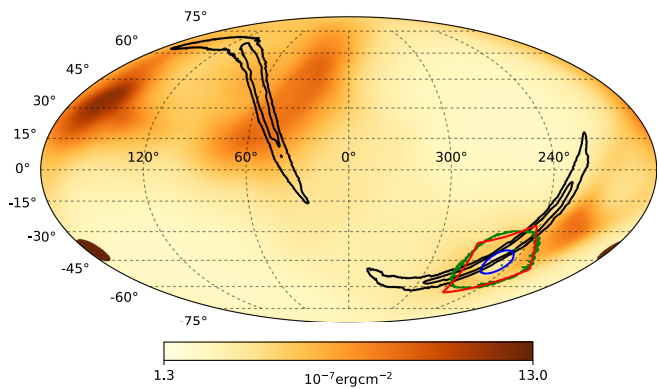}}
\else
  \includegraphics[trim={0cm 0cm 0cm 0cm},width=1.\columnwidth, angle=0]{sky_sens__1s.eps}
\fi
  \caption{Combined 3~$\sigma$ upper limit on the non-detection of the
    electromagnetic counterpart of \lvtevent derived by using the all
    sky observations of SPI-ACS, IBIS/Veto, IBIS/PICsIT and
    IBIS/ISGRI.  The color map indicates the upper limit values in
    different sky regions.  We assumed here the case of a 1~s
    long-lasting impulsive event with a spectral energy distribution
    described by the Comptonized model introduced in
    Sect.~\ref{sec:observations}. The model parameters are
    $\alpha=$-0.5 and
    $E_{peak}=$600~keV.  Black
    contours represent the most accurate localization of \lvtevent
    obtained from the LALInference \citep{triggerempaper} at both the
    50\% and 90\% confidence level. Red, green, and blue contours
    indicate the region covered by the INTEGRAL imaging FoV
    observations performed with IBIS/ISGRI, SPI, and the two JEM-X
    modules, respectively (the larger bounds of all these FoVs has
    been set to the point where a worsening of the instrument
    sensitivity by a factor of 20 compared to the on-axis value is
    reached), see also Figure~\ref{fig:sens_containment_soft}. During
    this observation the Earth shadow was near its maximal size for
    INTEGRAL (about 10~deg diameter) and did not occult the region of
    non-negligible LIGO probability.}

  \label{fig:sky_sens_total_1s}
\end{figure}

\ifpaper
\end{landscape}
\fi

\ifpaper
\newgeometry{left=1cm}
\begin{landscape}
\fi

\begin{figure}
  \centering 
\ifpaper
  \resizebox{\hsize}{!}{\includegraphics[trim={1cm 1cm 1cm 1cm},width=0.8\columnwidth, angle=0]{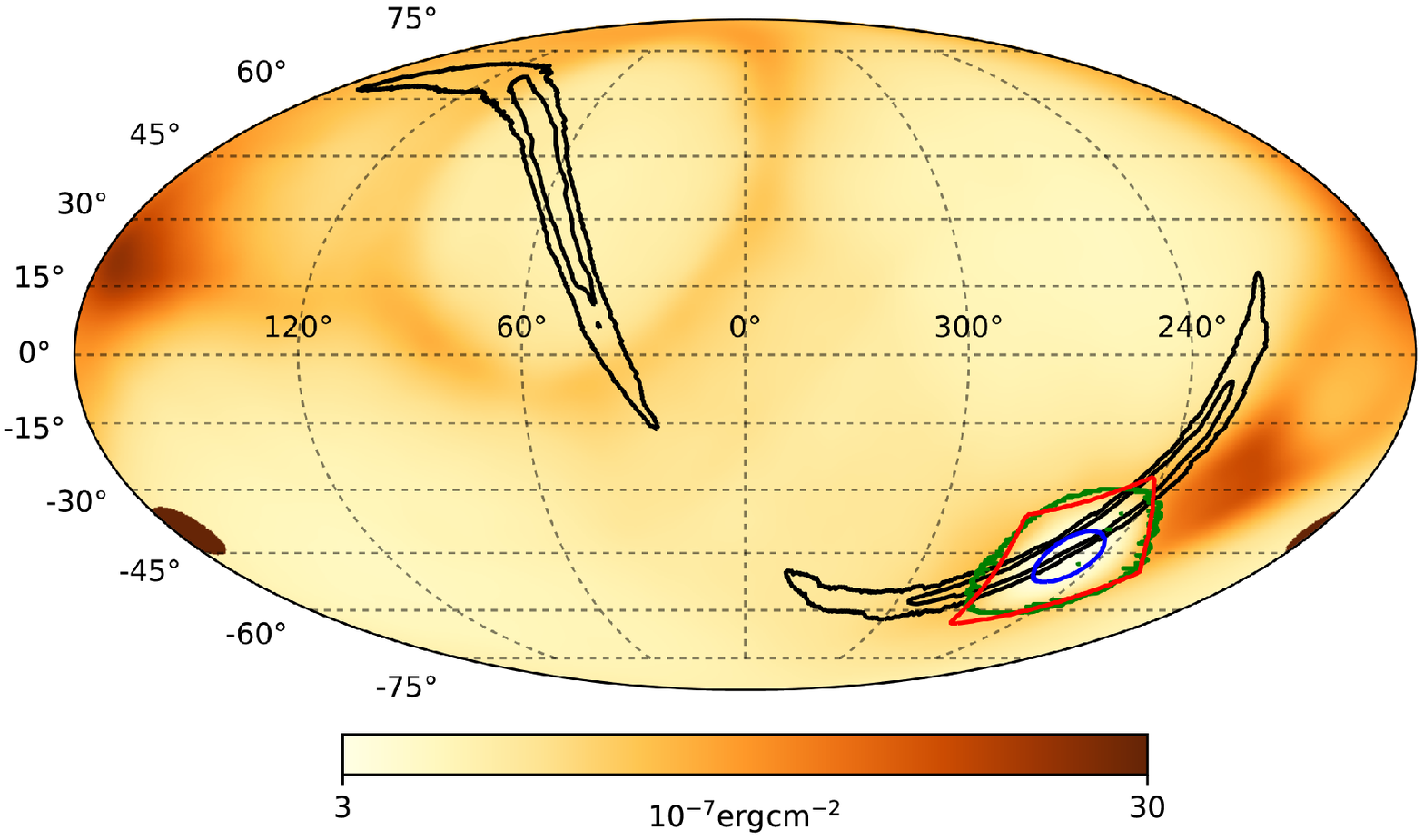}}
\else
  \includegraphics[trim={0cm 0cm 0cm 0cm},width=1.\columnwidth, angle=0]{sky_sens__8s.eps}
\fi
  \caption{Same as Figure~\ref{fig:sky_sens_total_1s} but in the case of a 8~s long-lasting impulsive event 
  with a spectral energy distribution described by the Band model with parameters $\alpha=$-1, 
  $\beta=$-2.5, and $E_{peak}=$300~keV, see also Figure~\ref{fig:sens_containment_soft}.}
  \label{fig:sky_sens_total_8s}
\end{figure}

\ifpaper
\end{landscape}
\fi

\ifpaper
\begin{landscape}
\fi

\begin{table*}[t]

\def\arraystretch{1.1}
\setlength{\tabcolsep}{0em} 
\footnotesize
\centering
\caption{Summary of sensitivities for the different instruments on-board INTEGRAL, related to the 
detection of an impulsive event as the counterpart of \lvtevent. }

\begin{tabular}{ c c@{\hskip-3pt} c@{\hskip-7pt} c@{\hskip-7pt} c@{\hskip-5pt} c@{\hskip-3pt} c c }

\toprule
 Instrument & energy range & \multicolumn{4}{c}{3$\sigma$ sensitivity}
\\ & & $\theta >$~120~deg & 120~deg~$ > \theta > $~80~deg & 80~deg~$ > \theta > $FoV & Field of View & On-Axis\\
 & & (erg cm$^{-2}$) & (erg cm$^{-2}$) & (erg cm$^{-2}$) & (erg cm$^{-2}$) & (erg cm$^{-2}$)  \\
\vspace{0.2cm}
\\
\midrule

JEM-X & 3~--~30~keV & \
 \ifdiff{N/A}{N/A} & \
 \ifdiff{N/A}{N/A} & \
 \ifdiff{N/A}{N/A} & \
 \ifdiff{2.6$\times$10$^{-7}$}{9.2$\times$10$^{-7}$} & \
 \ifdiff{4.5$\times$10$^{-8}$}{1.6$\times$10$^{-7}$} \\

ISGRI & 20~--~200~keV & \
 \ifdiff{N/A}{N/A} & \
 \ifdiff{N/A}{N/A} & \
 \ifdiff{2.3$\times$10$^{-7}$}{1.1$\times$10$^{-6}$} & \
 \ifdiff{8.3$\times$10$^{-8}$}{1.8$\times$10$^{-7}$} & \
 \ifdiff{3.5$\times$10$^{-8}$}{7.5$\times$10$^{-8}$} \\

PICsIT & 260~--~2600~keV & \
 \ifdiff{N/A}{N/A} & \
 \ifdiff{N/A}{N/A} & \
 \ifdiff{8.6$\times$10$^{-7}$}{2.4$\times$10$^{-6}$} & \
 \ifdiff{9.7$\times$10$^{-7}$}{1.9$\times$10$^{-6}$} & \
 \ifdiff{4.4$\times$10$^{-7}$}{8.7$\times$10$^{-7}$} \\

SPI & 25~--~8000~keV & \
 \ifdiff{N/A}{N/A} & \
 \ifdiff{N/A}{N/A} & \
 \ifdiff{N/A}{N/A} & \
 \ifdiff{1.6$\times$10$^{-6}$}{2.3$\times$10$^{-6}$} & \
 \ifdiff{3.9$\times$10$^{-7}$}{5.7$\times$10$^{-7}$} \\

SPI-ACS & 75~--~2000~keV & \
 \ifdiff{3.6$\times$10$^{-7}$}{10$^{-6}$} & \
 \ifdiff{2.2$\times$10$^{-7}$}{6$\times$10$^{-7}$} & \
 \ifdiff{2$\times$10$^{-7}$}{5.3$\times$10$^{-7}$} & \
 \ifdiff{3.5$\times$10$^{-7}$}{9.5$\times$10$^{-7}$} & \
 \ifdiff{4.2$\times$10$^{-7}$}{1.4$\times$10$^{-6}$} \\

IBIS/Veto & 100~--~2000~keV & \
 \ifdiff{8.7$\times$10$^{-7}$}{7.6$\times$10$^{-7}$} & \
 \ifdiff{2.1$\times$10$^{-6}$}{1.9$\times$10$^{-6}$} & \
 \ifdiff{1.7$\times$10$^{-6}$}{1.7$\times$10$^{-6}$} & \
 \ifdiff{2.7$\times$10$^{-6}$}{2.8$\times$10$^{-6}$} & \
 \ifdiff{N/A}{N/A} \\

\midrule
\multicolumn{2}{c}{fraction of the} &  48\% & 12\%  & 19~--~36\%  & 4~--~19\%\\
 \multicolumn{2}{c}{LVT151012 localization}   & \multicolumn{3}{c}{} \\
  &   & \multicolumn{3}{c}{} \\

\bottomrule
\vspace{0.3cm}
\end{tabular}

\label{tab:prompt}
\tablefoot{Average 3-sigma sensitivity for prompt events in the different regions of the sky, assuming two types of the transients:
  short-hard and long-soft; respectively left and right
  numbers. Short-hard burst is assumed to be 1~s long and
  characterized by a typical hard GRB spectrum, characterized by a
  Comptonized model (a parameterization of cutoff powerlaw power,
  see \cite{gruber14}) with parameters
  $\alpha=$-0.5,
  $E_{peak}=$600$~$keV. The long burst
  has a duration of 8~s and a typical soft GRB spectrum, i.e. a
  smoothly broken power law (Band model \citep{band93}) with
  parameters $\alpha=$-1,
  $\beta=$-2.5,
  $E_{peak}=$300$~$keV. }
\end{table*}

\ifpaper
\end{landscape}
\restoregeometry
\fi

\ifpaper
\begin{landscape}
\fi

\begin{table*}[t]
\label{tab:fov}
\centering
\caption{Summary of  sensitivities for the different instruments on-board INTEGRAL, related to the 
detection within their FoVs of a GRB afterglow-like event as the counterpart of \lvtevent.\ }
\begin{tabular}{ c c c c c c }

\toprule
Instrument & Energy range & Field of View  & Fraction of the \lvtevent\ &
Angular resolution &  3~$\sigma$ sensitivity \\
& & deg$^2$ & localization & & (erg~cm$^{-2}$~s$^{-1}$)  \\
\\
\midrule

JEM-X & 3~--~30~keV & \
 110 & \
 4~\% & \
 3' & \
 1.4$\times$10$^{-9}$ \\

ISGRI & 20~--~200~keV & \
 820 & \
 19~\% & \
 12' & \
 7.4$\times$10$^{-10}$ \\

PICsIT & 260~--~2600~keV & \
 820 & \
 19~\% & \
 30' & \
 2.4$\times$10$^{-8}$ \\

SPI & 25~--~1000~keV & \
 790 & \
 18~\% & \
 2.5$^\circ$ & \
 4.2$\times$10$^{-9}$ \\

\bottomrule
\vspace{0.3cm}
\end{tabular}
\label{tab:fov_longlasting}
\tablefoot{Data were 
integrated within the time interval spanning from a minimum of
T$_0$-30 to a maximum of T$_0$+287 (depending on the instrument) and a
power-law shaped spectral energy distribution with a photon index of
$-2$ has been assumed for the event to be detected. The limit of the
FoVs has been set to the point where a worsening of the instrument
sensitivity by a factor of 20 compared to the on-axis value is
reached.}
\end{table*}
\ifpaper
\end{landscape}
\fi

\section{Results}
\label{sec:results}

We combine in this section all previous results in order to derive the tightest possible upper limits 
on any electromagnetic 
emission in the entire LIGO localization region that can be associated with \lvtevent, depending on the 
assumed spectral energy 
distribution and event duration. 

\subsection{Impulsive events}
\label{sec:shortevent}

By exploiting the coverage of all instruments on-board INTEGRAL, we
could search for an impulsive event associated to \lvtevent in a wide
energy band, covering from 3~keV to 10~MeV. The most stringent upper
limits on the detection of such event were obtained in those regions
of the sky included within the imaging instruments FoVs, containing up
to about 20\% of the total LIGO localization probability of \lvtevent.

For the remaining regions, we could only derive upper limits on the
fluence of non-detected events above $\sim$75~keV by exploiting the
INTEGRAL omni-directional instruments and the out-of-FoV coverage
provided by IBIS/ISGRI and IBIS/PICsIT.  An effective way of
illustrating the relative capabilities of all INTEGRAL instruments for
these latter observations is provided in
Figure~\ref{fig:isens_zenith_hard}. We computed in these two figures the
range of detection significances achieved by each instrument for two
representative GRB-like events: a short burst (1~s) characterized by a
hard emission (Figure~\ref{fig:isens_zenith_hard} upper panel) and a
long burst (8~s) with a softer spectral energy distribution
(Figure~\ref{fig:isens_zenith_hard} middle panel; see also our previous
definitions in Sect.~\ref{sec:observations}). We assumed an arbitrary
fluence for the 1~s or 8~s-long event and computed for each zenith
angle the distribution of detection significances that can be achieved
by the different instruments at all azimuthal angles. We normalized
all these values to the highest detection significance obtained with
the SPI-ACS and then used the derived distributions (excluding for
clarity the 10\% worst and 10\% best values) to draw a shaded region
per instrument representing the fraction of normalized detection
significance (from 0\% to 100\%) as a function of the zenith angle.

We note from Figure~\ref{fig:isens_zenith_hard} that the detection
significance strongly depends on the zenith (i.e., off-axis)
angle. Taking as a reference the connecting axis between SPI and IBIS,
we can define two relevant hemispheres to explain this dependence. For the
SPI-ACS, the rightmost shaded region is more extended due to the
presence of IBIS, while the leftmost region is narrower as this
corresponds to the unobstructed portion of the sky visible by the
instrument.  The regions derived in the figure for the IBIS/Veto, IBIS/ISGRI, and
IBIS/PICsIT are more regular compared to those of the SPI-ACS, as the
whole SPI assembly covers a relatively small solid angle, as seen by
IBIS. The nearly vertical red lines represent the rapid transition
between high sensitivity within the IBIS/ISGRI FoV and the directions
for which the passive shield of IBIS starts to be important.  The
different orientation of all shaded regions highlight the
complementarity of the different instruments to efficiently cover the whole sky.

Based on the above results and those obtained in Sect.~\ref{sec:observations}, we conclude that 
in the case of short (1~s) impulsive events outside the imaging instruments FoVs, the most 
stringent upper limits are derived from the SPI-ACS in about 99\% of the sky, ranging from 
\ec{1.3$\times$10$^{-7}$} to 
\ec{7.1$\times$10$^{-7}$}. The IBIS/Veto is
more sensitive than SPI-ACS to these events in about 1\% of the sky.
For impulsive events lasting more than 8~s, IBIS/Veto is more
sensitive than the SPI-ACS in about 15\% of the sky (see
Figure~\ref{fig:sky_sens_4}). The sky fraction in which IBIS/Veto has a
high detection efficiency is much reduced compared to that of SPI-ACS,
but the upper limit on the non-detection of any electromagnetic
emission associated with \lvtevent that can be obtained from the
IBIS/Veto is a factor of 4 more stringent in those regions that are
only accessible with a reduced SPI-ACS sensitivity (at 3~$\sigma$
confidence level and for 8~s long events).  For events occurring
inside the FoV of different instruments, ISGRI provides the most
stringent upper limits for both short-hard and long-soft impulsive
events, while SPI is best suited to study particularly hard events
with most of the photons released at energies $\gtrsim$200~keV.  JEM-X
provides, in principle, coverage within the energy range 3--35~keV,
but its smaller FoV compared to ISGRI makes serendipitous detections
of impulsive events less likely.

\begin{figure}
  \centering \resizebox{\hsize}{!}{ \includegraphics[width=0.9\columnwidth]{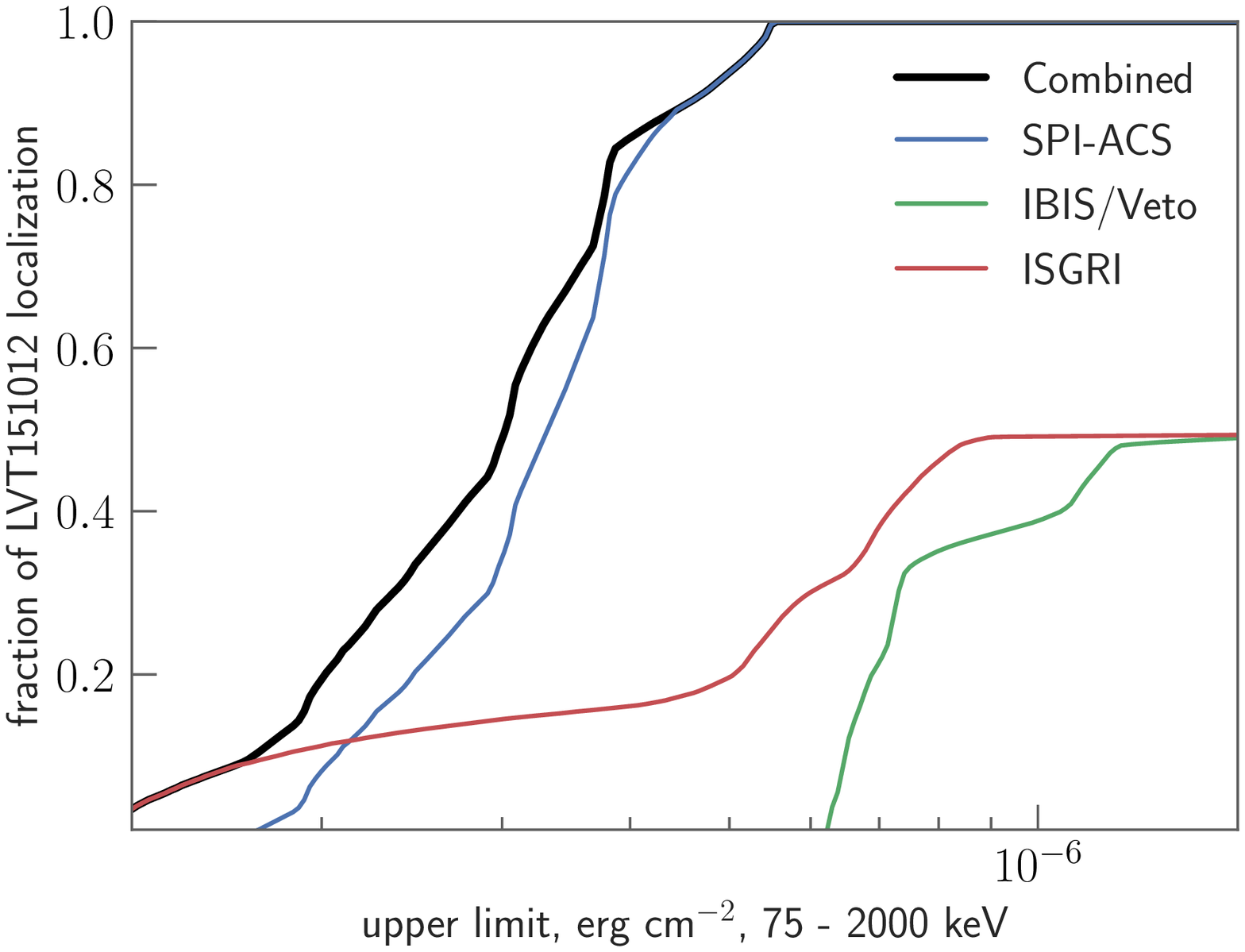}} \caption{\change{Probability
  of \lvtevent as function of combined and individual instrument
  3~$\sigma$ upper limits on the non-detection of its electromagnetic
  counterpart .  We assumed here the case of a 1~s long-lasting
  impulsive event with a spectral energy distribution described by the
  Comptonized model introduced in Sect.~\ref{sec:observations}. The
  model parameters are $\alpha=$-0.5 and
  $E_{peak}=$600~keV.
}  }  \label{fig:sens_containment_hard}
\end{figure}

\begin{figure}
  \centering \resizebox{\hsize}{!}{ \includegraphics[width=0.9\columnwidth]{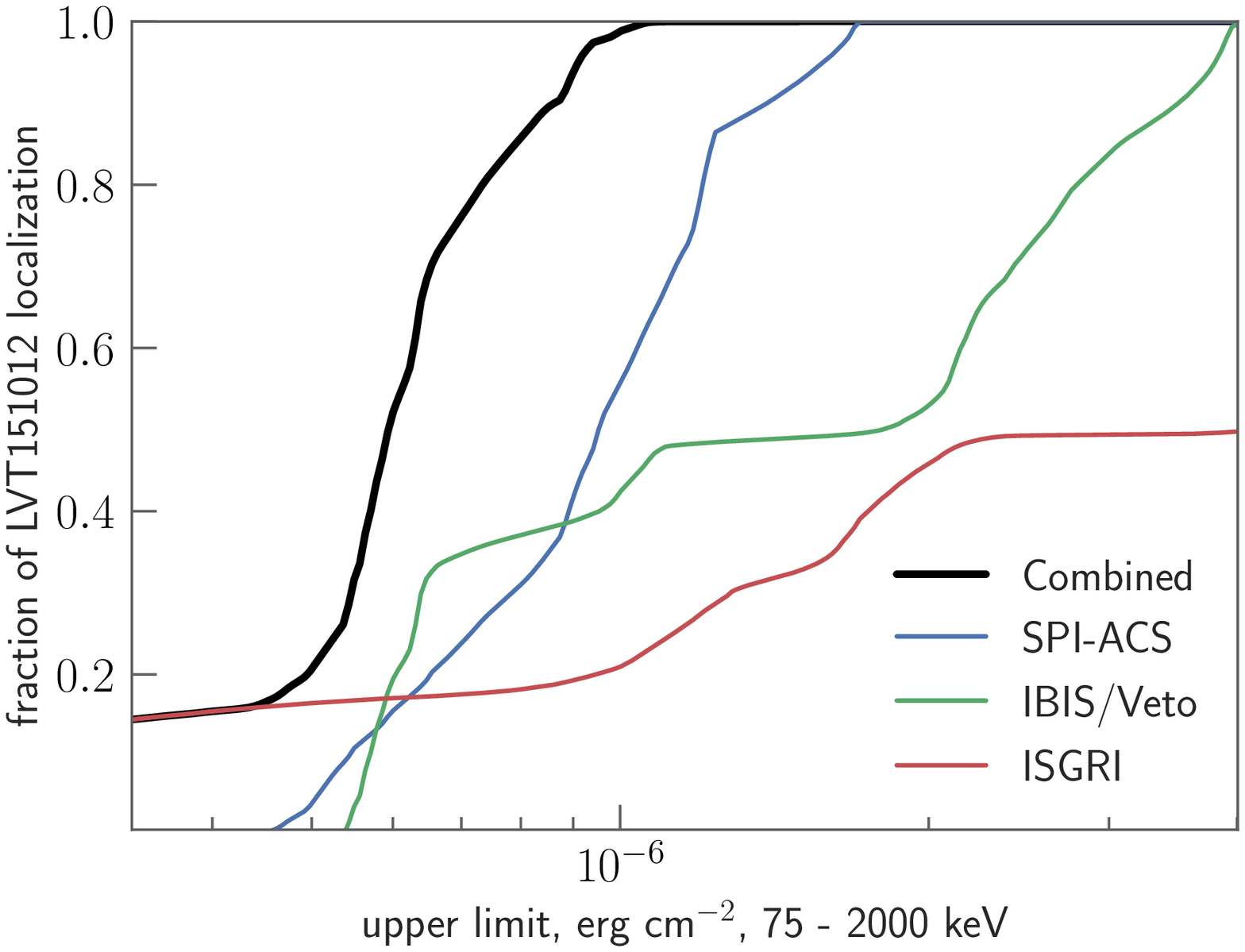}} \caption{Same
  as Figure~\ref{fig:sens_containment_hard} but in the case of a 8~s
  long-lasting impulsive event with a spectral energy distribution
  described by the Band model with parameters
  $\alpha=$-1,
  $\beta=$-2.5, and
  $E_{peak}=$300~keV.}  \label{fig:sens_containment_soft}
\end{figure}

\change{The capability of INTEGRAL to constrain gamma-ray emission from the entire
highly extended LIGO localization region can be efficiently illustrated
by computing the LVT151012 localization probability observed with
different levels of sensitivity, see
Figures~\ref{fig:sens_containment_hard}
and \ref{fig:sens_containment_soft}.}

Summarizing, we can identify three zones within the LIGO localization area of \lvtevent in which 
INTEGRAL can provide stringent upper limits on the fluence of its possible electromagnetic counterpart and which 
are not included within the FoVs of INTEGRAL imaging instruments:  

\begin{enumerate}
\item off-axis angles up to 80~deg (excluding the instruments FoVs). This region is simultaneously 
covered with almost the best achievable sensitivity of  SPI-ACS and IBIS/ISGRI+PICsIT for out-of-FoV observations.
\item off-axis angles ranging from 80 to 120~deg. This region is covered with the best achievable sensitivity for SPI-ACS. 
\item off-axis angles larger than 120 deg. In this case, SPI-ACS can only cover the region with limited sensitivity, but 
the IBIS/Veto almost achieves its best performances. 
\end{enumerate}

The sensitivity of different instruments for the detection of a 1~s or 8~s long-lasting event in these regions 
is summarized in Table~\ref{tab:prompt} and Figure~\ref{fig:spec_sens_fov_short}.

\subsection{GRB afterglow-like emission}
\label{sec:afterglow} 

The upper limits on the non-detection of a GRB afterglow-like emission 
($F_E \sim E^{-2}$) from \lvtevent that we obtained inspecting the FoV of  imaging instruments 
on-board INTEGRAL (JEM-X, IBIS/ISGRI, IBIS/PICsIT, and SPI) are summarized in Table~\ref{tab:fov}. 
ISGRI provides the highest sensitivity to search for such emissions 
within the IBIS FoV (see Figure~\ref{fig:sens_4fov}). Harder events with the bulk of  
high energy radiation emitted above $\sim$200~keV can be more efficiently detected by  
SPI. JEM-X is the only instrument providing coverage at energies $<$20 keV.   

These results are particularly relevant for the dedicated target of opportunity (ToO) 
observations that are being planned with INTEGRAL in order to follow-up future triggers 
provided by LIGO and Virgo. These ToO observations will image the accessible regions of the sky 
included in the localization of  GW events within $\sim$1~day from the 
trigger, thus probing mainly the presence of any long-lasting electromagnetic emission 
from the GW source. 
The complementary of INTEGRAL instruments provides a unique broad-band coverage 
and is very well suited to hunt for electromagnetic counterparts of GW events.

\section{Discussion}
\label{sec:discussion}

INTEGRAL features a duty cycle as high as 85\% for the collection of
scientific data, being unable to monitor the X-ray/$\gamma$-ray 
sky only for a short amount of time around the perigee passage when
all instruments are turned off due to charged particles in the Earth
radiation belts.  Its elongated orbit strongly limits the occultation
of the accessible sky by Earth, making the suite of all available
on-board instruments capable to efficiently catch the prompt emission
associated with GW events, as well as to follow up any possible
related afterglow in a broad energy range (3~keV--10~MeV).

SPI-ACS is the most sensitive all-sky detector of $\gamma$-rays with
energies $\gtrsim$100~keV among the instruments currently in operation
or solidly planned in the foreseeable future.  It can be efficiently
exploited to hunt for the prompt emission associated to LIGO/Virgo
triggers, and could provide in case of \lvtevent relatively deep
3~$\sigma$ upper limits on the fluence of non detected counterparts on
the all sky, ranging
from \ec{1.3$\times$10$^{-7}$}
to \ec{2.2$\times$10$^{-6}$} (see
Table~\ref{tab:prompt} and Figure~\ref{fig:sky_sens_4}).  IBIS
(IBIS/Veto, IBIS/ISGRI, and IBIS/PICsIT) is also able to provide
monitoring of a large fraction of the sky, efficiently complementing
SPI-ACS data where its sensitivity is reduced. Moreover, we
showed that for the majority of transients more than one INTEGRAL
all-sky instrument is able to provide significant detection, opening a
possibility of constraining the signal arrival direction and its
nature. Even with the current level of systematic uncertainty, in some
particularly favorable cases, it is possible to narrow down to
uncertainty on the event location to within 1\% of the sky.

\change{Combination of all-sky INTEGRAL detectors allows to put stringent
upper limit on the hard X-ray emission associated with \lvtevent. This
limit can be interpreted as a constraint on ratio of isotropic
equivalent electromagnetic energy emitted in an isolated burst during
the \lvtevent to the total energy of the gravitational waves:
$E_{75-2000~keV}/E_{GW}$<4.4$\times$10$^{-5}$
$\left[\frac{D}{1000~Mpc}\right]^2
\left[\frac{F_{75-2000~keV}}{10^{-6}~erg~cm^{-2}}\right] \left[\frac{E_{GW}}{1.5\times M_{\sun}c^2}\right]^{-1}$. This constraint is less tight than the one we found in the case of GW150914 \citep{savchenko16} for several reasons: the LIGO event was more distant, less luminous, localized less favorably for SPI-ACS and took place near the end of INTEGRAL observations when the background was somewhat disturbed.}

As the FoV of IBIS and SPI corresponds to only 2\% of the sky, events
similar to the prompt emission of GRBs are rarely observed within
these limited areas (the situation is even worst for the two JEM-X
modules, given the much smaller FoV compared to both IBIS and SPI).
Although the number of correspondingly significant reported detections
are relatively small \citep[about ten detections a year for
IBIS;][]{Mereghetti2012}, the high INTEGRAL sensitivity in the hard
X-ray domain efficiently probes a population of particularly faint and
soft $\gamma$-ray bursts, with 20--200~keV fluences down to a few
10$^{-8}$~erg~cm$^{-2}$ \citep{Bosnjak2014}.  If at least a fraction
of the LIGO/Virgo localization area for the future GW triggers will
overlap with the IBIS and/or SPI FoV, INTEGRAL observations could be
used to localize and characterize in fairly good details the
correspondingly electromagnetic counterparts, or at least put tight
constraints on their hard X-ray/$\gamma$-ray emission (similarly to
what we have done in the present paper for \lvtevent).  The
possibility of exploiting the presence of the coded masks to obtain
the JEM-X, IBIS, and SPI images allows us also to perform in these
cases searches for any possible GRB afterglow-like emission associated
with the GW event.  This will be the main goal of the deep pointed
observations that INTEGRAL aims at performing within $\sim$1~d from
the announcement of future LIGO/Virgo triggers, whenever compatible
with the satellite operational constraints.

\change{For the fraction of the \lvtevent probability region in the field of
view of the INTEGRAL detectors we are able to put tight upper limit on
the long-lasting hard X-ray emission component. This implies a
constrain on ratio of isotropic equivalent electromagnetic energy
emitted in the 287~s following the \lvtevent to total energy of the
gravitational waves:
$E_{75-2000~keV}/E_{GW}$<8.9$\times$10$^{-6}$
$\left[\frac{D}{1000~Mpc}\right]^2
\left[\frac{F_{20-200~keV}}{2\times 10^{-7}~erg~cm^{-2}}\right] \left[\frac{E_{GW}}{1.5\times M_{\sun}c^2}\right]^{-1}$. Here we have chosen energy range the most suitable for IBIS/ISGRI, which is the most sensitive instrument to a long-lasting emission characterized by our assumed spectral energy distribution  (a powerlaw with a slope of $-2$).}

The nature of the hard X-ray emission that we should expect from the
likely associated with GW events remains speculative.  Afterglows of long and
short GRBs are amongst the most anticipated counterparts of the LIGO
and Virgo detections, but their hard X-ray properties are not yet well
understood.  When a bright nearby long GRB is observed by a sensitive
hard X-ray instrument, it may remain visible for many hours.  Two
examples are those of GRB\,990123, which was detected at energies
$\gtrsim$20~keV by BeppoSAX/PDS up to 20~ks after the onset of the
event \citep{Corsi2005}, and GRB120711A, which was observed by
INTEGRAL/ISGRI (INTEGRAL/JEM-X) to faint down not earlier than
10~ks \citep[20~ks;][]{MartinCarrillo2014}.  The long-lasting hard
X-ray emission of GRB120711A was shown to be remarkably similar to
that recorded in the GeV band by the Fermi/LAT. As the detection of
luminous GRB afterglows are not uncommon at these very high
energies \citep{ackermann2013}, it seems reasonable to expect that an
associated hard X-ray emission should also be frequently
observed. Convincing evidences for the presence of a slowly decaying
hard X-ray and soft $\gamma$-ray afterglow emission has been reported
for several GRBs, such as GRB\,130427A \citep{maselli130427a} and
GRB\,080319B \citep[,Figure 4.14]{Savchenko2012a}. In the case of
GRB\,130427A, the connection between the afterglow in the hard X-ray
and $\gamma$-ray bands was supported by a highly sensitive pointed
NuSTAR observation
\citep{Kouveliotou2013}. The paucity of previous detections at hard X-rays could be mostly 
ascribed to the relatively narrow FoV of the instruments available in this energy 
domain and the lack of performed follow-up observations of GRB afterglows.  

Rescaling the hard X-ray afterglow of GRB~120711A (redshift of 1.4 or
D$_L$ = 10~Gpc) to the distance of LVT151012 that also corresponds to
the typical observational horizon of the LIGO O1/O2 runs (redshift of
0.2 or D$_L$= 1~Gpc), would make such an event significantly
detectable in ISGRI and JEM-X in the 10~days following the initial
onset of the burst (we assumed here that the decay in time continued
accordingly to what was observed during the initial afterglow of
GRB~120711A or, in the worst case, that it followed a power-law shaped
decay in time characterized by a slope not steeper than $t^{-2}$). In
case the direction of an outflow from a GRB is not aligned with the
observer's line of sight, the high energy luminosity during an
afterglow will be suppressed and will reach its peak at a later
time \citep{Granot2002}.  Comparing the 3-$\sigma$ sensitivity of
INTEGRAL/ISGRI to the luminosity of the afterglow recorded from
GRB~120711A, rescaled to a distance of 1~Gpc, we found that ISGRI
could have detected such event even if the outflow was not originally
pointed towards the observer, as long as the angle between the outflow
direction and the line of sight was no more than 30~degrees.  Note
that, assuming a values of a GRB jet opening angle similar to that
measured in the case of GRB~120711A, only few percent of the
misaligned detections would be accompanied by a prompt emission. This
suggests the presence of a considerable population of orphan hard
X-ray afterglows. A similar result can be obtained for the case of
GRB~130427A using the outcomes of dedicated multiwavelength
observations.  The hard X-ray emission associated to the afterglow of
this exceptionally bright burst was observed by Swift/BAT and NuSTAR,
which operate in low-Earth orbits and can point continuously to a
celestial target for no longer than $\sim$3\,ks. As the highly
elliptical orbit of INTEGRAL enables uninterrupted observations with
duration up to about two days, we estimated that the afterglow of
GRB\,130427A should have remained detectable to the INTEGRAL
instruments for about 1 day following the onset of the event.

In at least a few cases, the prompt emission of the GRB was observed
to last for more than 20~ks \citep[e.g.][]{Evans2014,
grb130925a_gcn}, and thus a very rapid hard X-ray follow-up could
catch the burst still going off while giving rise to an associated GW
emission.

Even though the above arguments show that there are promising
prospects to observe the afterglows of long GRBs with INTEGRAL, it is
presently expected that GW events should be associated to short GRBs
following the merging of compact binary systems involving at least a
neutron star. So far, no hard-X-ray afterglows have been detected from
short GRBs. Fermi/LAT observations of GRB\,090510 provided some
indications that particularly bright short GRB could display a
$\gamma$-ray afterglow. We conclude that the presence of an hard X-ray
afterglow following at least a few short GRBs cannot be completely
ruled out yet.

On the other hand, it is also possible that energetic supernovae,
accompanied by a bright and long GRB, produce a distinct gravitational
wave signal \citep[see][for a review]{Ott2008}.

Magnetars are also suspected to produce GWs with sufficient intensity
to be detected by Advanced LIGO and Advanced Virgo
\citep[e.g.][]{Zink2011}. Several objects in this class showed 
outbursts lasting for a few days during which many bright flares of
hard X-rays were going off, with a clearly detectable high energy
emission extending up to 100~keV, producing remarkable signal in the
INTEGRAL detectors \citep{mereghetti09,savchenko10}. In the future, GW
signals possibly associated with these events could be efficiently
followed-up with INTEGRAL.

Future GW detections might also pave the way for still unexplored (or
poorly explored) physical mechanisms to trigger substantial
electromagnetic emissions during these events.  This was already the
case when the tentative gamma-ray counterpart of GW150914 was
announced \citep{gbmpaper}. A BH-BH merger is expected to have little
(if any at all) material in its immediate surroundings and was thus
not considered a likely source of electromagnetic emissions so
far. However, it has been speculated that even in this case a GRB-like
event may be expected under some peculiar assumptions, such as the
presence of a long-lived accretion disk \citep{Perna2016}, the
coalescence immediately following the collapse of a star
\citep{Loeb2016,Woosley2016}, the non-vanishing electric charge of a 
black holes \citep{Fraschetti2016} or the creation of a transient naked
singularity \citep{Malafarina2016}. If a binary BH merger can be associated with beamed
relativistic outflows, their interaction with the close-by ambient
medium might led to the production of a long-lasting electromagnetic
emission observable from the radio to TeV energies, similarly to what happens 
during a GRB afterglow \citep{Zhu2016}.

\section{Conclusions}
\label{sec:conslusions}

At odds with the cases of GW150914 and GW151226, there was no
extensive multi-wavelength campaign to follow-up the discovery
of \lvtevent\ due to the late release of the corresponding
announcement by the LIGO/Virgo collaboration. Luckily, INTEGRAL was
pointing towards the peak of the LIGO localization probability,
allowing us to exploit the full capabilities of its instruments.  We
reported in this paper the serendipitous search for electromagnetic
counterparts to the possible GW event by exploiting all INTEGRAL data
that were collected at that time.  The results of our analysis provide
the most sensitive constraints obtained so far on the non-detection of
any electromagnetic emission associated to \lvtevent\ in a large
fraction of the event localization, as well as very competitive limits
in the whole 90\% localization region. This confirms and strengthens
the potential of INTEGRAL to hunt for GW counterparts, as already
demonstrated at the time of GW150914.

Using the SPI-ACS, we set a 3~$\sigma$ upper limit on the whole
localization region of \lvtevent\ ranging
from \ec{1.3$\times$10$^{-7}$}
to \ec{2.2$\times$10$^{-6}$}
in 75~--~2000~keV (depending on the source location,
spectrum, and duration). This result rules out the possibility that a
bright impulsive burst of $\gamma$-rays is associated with the GW
\change{event, constraining the ratio of the isotropic equivalent energy
released in such an event to the energy of the GWs:
$E_{75-2000~keV}/E_{GW}<$ 4.4$\times$10$^{-5}$}.
Additional upper limits are provided for specific limited regions in
the sky, taking advantage of the complimentary coverage provided by
the second omni-directional instrument on-board INTEGRAL, the
IBIS/Veto, and the two imaging instruments capable of providing
coverage also outside their FoV, i.e. the IBIS/ISGRI and
IBIS/PICsIT. In particular, the data of IBIS/Veto reach a better
sensitivity compared to those of SPI-ACS in the sky region located at
off-axis angles larger than $120$~deg with respect to the satellite
aim point, while ISGRI and PICsIT improve the sensitivity for
locations in the sky up to 15~deg off-axis (30~deg in case of events
characterized by a particularly soft spectral energy distribution).
The most stringent upper limits for impulsive events associated
with \lvtevent were obtained within the FoV of the coded-mask imaging
instruments IBIS, SPI, and JEM-X, which overlapped with a large
portion of the higher probability localization region of the GW event.
We set a 3~$\sigma$ upper limit of
\ec{3.5$\times$10$^{-8}$} (\ec{7.5$\times$10$^{-8}$}) in the 20~--~200~keV energy range, 
\ec{3.9$\times$10$^{-7}$} (\ec{5.7$\times$10$^{-7}$}) in the 25~--~8000~keV energy range, and
\ec{4.5$\times$10$^{-8}$} (\ec{1.6$\times$10$^{-7}$}) in the 3~--~30~keV energy range,  in the case of a short-hard (long-soft) GRBs-like impulsive event observed close to the spacecraft aim point. 
All these results are summarized in Table~\ref{tab:prompt}.

The availability of images from IBIS, SPI, and JEM-X allowed us also
to search in the correspondingly covered fraction of the sky for GRB
afterglow-like events possibly associated with \lvtevent. Also in this
case no significant detection was found, \change{limiting the ratio of the
isotropically equivalent energy in slowly decaying hard X-ray
afterglow to total GW energy to
$E_{20-200~keV}/E_{GW}<$ 8.9$\times$10$^{-6}$}.  We
summarized all results in Table~\ref{tab:fov_longlasting}.  For future
triggers, that will be promptly announced by the LIGO/Virgo
collaboration, INTEGRAL plans dedicated pointed observations that will
cover as much as possible of the event localization regions with the
imaging instruments to search and characterize long-lasting high
energy emission associated with GW events.  The joint effort of Virgo
and LIGO might also lead in the future to an enhanced localization
capability for GW events that could allow INTEGRAL to cover the entire
region with a single pointing.

Our non-detections of any electromagnetic counterpart to \lvtevent 
(to within the limiting observational capabilities of the instruments on-board INTEGRAL)  
remains compatible so far with the commonly accepted scenario that no hard X-ray/$\gamma$-ray emission is expected 
from the coalescence of a binary black hole.  

\begin{acknowledgements}
This paper is based on data from observations with INTEGRAL, an ESA
project with instruments and science data center funded by ESA member
states (especially the PI countries: Denmark, France, Germany, Italy,
Spain, and Switzerland), Czech Republic and Poland, and with the
participation of Russia and the USA.  The SPI-ACS detector system has
been provided by MPE Garching/Germany. We acknowledge the German
INTEGRAL support through DLR grant 50 OG 1101. The Italian
INTEGRAL/IBIS team acknowledges the support of ASI/INAF agreement
n. 2016-025-R.0. AL and RS acknowledge the support from the Russian
Science Foundation (grant 14-22-00271). Some of the results in this
paper have been derived using the \software{HEALPix} \citep{healpix}
package. We are grateful the Fran\c cois Arago Centre at APC for
providing computing resources, and VirtualData from LABEX P2IO for
enabling access to the StratusLab academic cloud. Finally we thank the
referee for careful reading of the manuscript and for the insightful
comments.

\end{acknowledgements}

\appendix
\section{Instrument description}

\subsection{JEM-X} \label{sec:jemx}

JEM-X is a coded-mask X-ray monitor consisting of two identical Microstrip Gas Chamber
modules which are simultaneously operational since 2010 \citep{jemx}.  
The two modules are surrounded by a collimator tube that is nearly opaque to X-rays between  
3 and 35~keV, where JEM-X is collecting science data. For this reason, the instrument 
is not useful to search for impulsive events occurring outside its FoV. 
As JEM-X is endowed with a relatively large FoV, featuring a diameter of 13.2~deg at zero response, and 
provides coverage in an energy range that is not accessible by any other INTEGRAL instrument, it is 
particularly well suited and best exploited to search for long-lasting soft X-ray transients. 
The extraction of reliable measurements from the JEM-X science data is not always straightforward, as the
response of the detector is occasionally unstable during the observations, owing
to temperature changes, irregularities introduced by
detector cold start, or autonomous switch-offs and reactivations.

\subsection{IBIS} \label{sec:ibis}

IBIS is a hard X-ray telescope characterized by a relatively good angular
resolution of about 12~arcmin and a point source location accuracy
(PSLA) of 1--3 arcmin (for sources with typical fluxes of a few mCrab in the 20-40~keV energy range). 
The reconstruction of  sky images is performed by exploiting the availability of a coded mask located 
3.2~m above the detection plane at the extreme end of the telescope. 
This results in a 9$\times$9 degree fully-coded FoV, extending
to about 30$\times$30 degrees when the zero-response partially coded FoV is considered.

The IBIS collimation system is made by using a passive shield. Its 
structural components are composed by a 1~mm thick tungsten hopper (placed on top of the
detection systems) and a lead tube with a variable thickness extended 
up to the mask. A 1.5~mm thick lead door complements the shield in the
direction opposite to SPI (see Figure~\ref{fig:isens_zenith_hard} bottom panel). 
The whole system, becoming increasingly transparent for energies above 200\,keV, was designed to reduce the
diffuse sky component of the instrument background.

The detection plane consists of 2 layers. The top, ISGRI, is optimized for the collection of photons up to 
energies of $\la 500$\,keV and placed 90~mm above the other layer, PICsIT. The latter covers a
higher energy range extending up to 10~MeV.  Both detector arrays are surrounded by an active
shield, the IBIS/Veto system, which is essential to reduce the cosmic-ray-induced instrument background.

\subsubsection{ISGRI} \label{sec:isgri}

IBIS/ISGRI \citep{theisgri} is the top detector layer of the IBIS
telescope and is made of 16384 4x4~mm and 2~mm thick CdTe pixels. It
is sensitive to photons with energies between 15 and 800\,keV and retains a detection 
efficiency of nearly 100\% for photons up to $\sim$100~keV. The effective area of ISGRI is 
reduced in the energy range 100~keV-1~MeV, as it becomes increasingly optically thin to such 
radiation. In this energy range, photons begin to interact with the full volume of the
detector pixels. The charge collection for interactions occurring at large depths in the detector is
less efficient, complicating reconstruction of the incident photon energy. Nevertheless, ISGRI maintains good 
energy resolution up to 800~keV, thanks to the original software-based charge loss correction. The
charge loss is reduced for photons arriving from highly off-axis
sources, thus increasing the instrument performances for the study of such objects. 
The instrumental energy resolution toward the end of 2015 ranged from 15\% at 60 keV to 9\% at 511~keV 
(full width half maximum, FWHM). These values slowly worsen with time due to the degradation of the detector 
gain. The latter is related to the evolution of  charge collection efficiency 
driven by the long term irradiation of the detector in space \citep{Caballero2013}, becoming the principal 
cause for the current 
 systematic uncertainties in the instrument calibrations.
For the same reason, the characteristic low-energy threshold of the instrument increased 
from 15~keV at launch to 22~keV in 2015, at the same time becoming progressively less sharp. Simultaneous 
increase and smoothening 
of low threshold efficiency cut-off implies that photons with energies not far above 15~keV 
can still be detected after more than 14~years in orbit, albeit with dramatically decreased efficiency.

The coded mask through which ISGRI is usually observing the high-energy sky cannot be fully exploited 
when searching for impulsive events within the fully-coded FoV, when their location is not known. While the
sensitivity for a source in a fixed location can be improved by using the coded mask pattern to reject 
about 50\% of the background, this advantage is lost in a search for a new source due to additional
trial factor. The conditions are different in the partially coded FoV, as a progressively smaller fraction of the detector is exposed 
through the coded mask holes and searches for short transients  could be optimized  by considering a smaller portion 
of the detector relevant for specific directions. 
This reduces the background, which is proportional to the 
total effective area used for the search. The instrument sensitivity 
is, however, also reduced by exploring lower effective areas, rapidly approaching that of 
the SPI-ACS (see Figs.~\ref{fig:sens_4fov} and \ref{fig:isens_zenith_hard} upper panel). 
We generally prefer to rely on the light curves built from the entire detector 
to search for impulsive events in the ISGRI data.   

As the IBIS collimator tube is becoming increasingly transparent at energies above $\sim$200~keV, photons  
from directions that are up to 80~deg off-axis with respect to the satellite aim point 
can reach the ISGRI detectors, making this instrument capable to detect also events occurring 
outside its FoV. Even soft events, with the bulk of  photons released below $\sim$200~keV, can also be 
detected in ISGRI despite the absorption by the IBIS shield and other 
satellite structures. For the very same reason, photons from soft events produce a highly contrasted 
pattern  
on the ISGRI detector plane, which can be used to roughly constrain the source location (similarly to what is done with
a higher precision when the event is recorded through the coded mask within the instrument FoV).  

ISGRI is particularly well suited to search for rather long transients, i.e.,
those associated to GRB afterglows, because the coded mask imaging allows us to 
better characterize the instrument background as well as persistent sources, accurately subtracting their 
contribution from the data 
to probe the presence of faint transients.

\subsection{PICsIT} \label{sec:picsit}

IBIS/PICsIT \citep{labanti03} is the bottom detector layer of the IBIS telescope and it is 
located 90 mm below ISGRI. It is made of 4096 30 mm-thick
CsI pixels, featuring a total collecting area of about
2900~cm$^2$ and sensitive to photons with energies between 175~keV
and 10~MeV. In this energy range, the IBIS collimator tube is
largely transparent and thus PICsIT can observe sources from all the directions
that are not occulted by IBIS/Veto (i.e. with off-axis angles as large as 70~deg; see
Sect.~\ref{sec:ibisveto}). The
effective area of the instrument is slowly decreasing as a function of off-axis angle,
mainly due to the effect of the PICsIT planar geometry combined with the
change in opacity of the shield and ISGRI detector plane. As the instrument coded mask opacity 
to hard X-ray photons is larger than that of the passive shield, 
PICsIT in principle collects more signal from sources outside the FoV rather than those closer to the 
satellite aim point (at odds with the case of ISGRI). This leads to a sensitivity increase for 
isolated bright impulsive events. 
In these cases, the background variability  can be often neglected and its average level well constrained 
before and after the event, without relying on the coded mask.

To evaluate the response of PICsIT to high energy bursts from any sky direction,
it is important to take into account the partial absorption of the corresponding 
radiation by the satellite structures. We thus performed Monte-Carlo 
simulations using the INTEGRAL mass model previously described by \citet{Ferguson2003} and  
improving it through the inclusion of a more detailed IBIS mass model \citep{laurent2003}. 
We validated our approach by comparing the results obtained for the detection of 
sources within the instrument FoV with the predictions of the PICsIT responses based on the
up-to-date instrument calibrations provided by the instrument team. 

\subsubsection{IBIS Compton mode} \label{sec:comptonmode}

When an incident photon undergoes Compton scattering in ISGRI, the
scattered photon may be subsequently detected by PICsIT. It is also
possible that a photon is Compton back-scattered in PICsIT and then
detected in ISGRI. The electronics of IBIS identifies such nearly
simultaneous interactions and records them in the so-called Compton
mode data \citep{ibis}. The effective area of the Compton mode is
lower than that of both ISGRI and PICsIT, reaching  
50~cm$^2$ at 400~keV for a 30~deg off-axis source. 

In principle, it is possible to use the Compton kinematics to
constrain the photon arrival direction and reject spurious
coincidences between independent ISGRI and PiCSIT events, being these
the main background component.  When searching for a known source, the
Compton-mode sensitivity can be improved by excluding events
incompatible with a given source location \citep{Forot2007}.  The
localization area for a single photon is primarily limited by the
PICsIT energy resolution and can be as small as 5\% of the sky for a
2~MeV incident photon. However, its size strongly depends on the
energies of incident and Compton scattered photons.  For a typical
background spectrum, only a small fraction of the interactions can be
well-localized: appropriate event selections provide an average
localization area per photon of about 20\% of the sky.  Selections
based on the event localization accuracy suppresses the background and
improves the determination of source coordinates, but reduces the
number of usable photons.  The trade-off between the event
localization accuracy and the number of usable events has to be
evaluated on a case-by-case basis. Our results on \lvtevent apply to a
wide range of conditions and should be considered representative in
the case of weak transient sources.

In the case of brighter events, the Compton mode offers an important
advantage, as it suffers much less telemetry stream saturation gaps
than both ISGRI and PiCSIT.  If the prompt emission associated with a
GW event is bright enough to be detected in the IBIS Compton mode
data, the localization of this event could be even more precise.  We
expect that an angular resolution better $\sim$5~deg can be achieved
for a 1~s-long event characterized by a spectral energy distribution
similar to that of a short-hard GRB and a fluence larger than
$5 \times 10^{-5}$~erg~cm$^{-2}$ (assuming the event occurs in the
area covered with the optimal sensitivity of the IBIS Compton mode). A
GRB-like event yielding a 20~$\sigma$ detection in the IBIS Compton
mode would also give rise to a detection at a significance of at least
50~$\sigma$ in SPI-ACS (300~$\sigma$ for most of the accessible source
directions).  We thus expect to exploit data collected by IBIS in
Compton mode only when a very strong detection of the same event is
reported in SPI-ACS, IBIS/ISGRI, and IBIS/PICsIT.

\change{The potential of the IBIS Compton mode to localize and 
characterize bright transients from at least 50\% of the the sky was
previously demonstrated by \cite{Marcinkowski06,isgrioffcat}, but was not
systematically exploited yet.}

\subsection{IBIS/Veto} \label{sec:ibisveto}

The bottom and lateral sides of the IBIS detectors are surrounded by an
active anti-coincidence shield, IBIS/Veto, which is made of 2~cm thick BGO
crystals \citep{Quadrini2003}. The count-rate of IBIS/Veto is continuously 
integrated every 8~s and transmitted to the ground, making this 
subsystem an efficient detector of GRBs (as well as other $\gamma$-ray transient phenomena), 
albeit with a reduced sensitivity for events shorter than the nominal integration time.

\begin{figure*}
  \centering
  \includegraphics[width=\columnwidth]{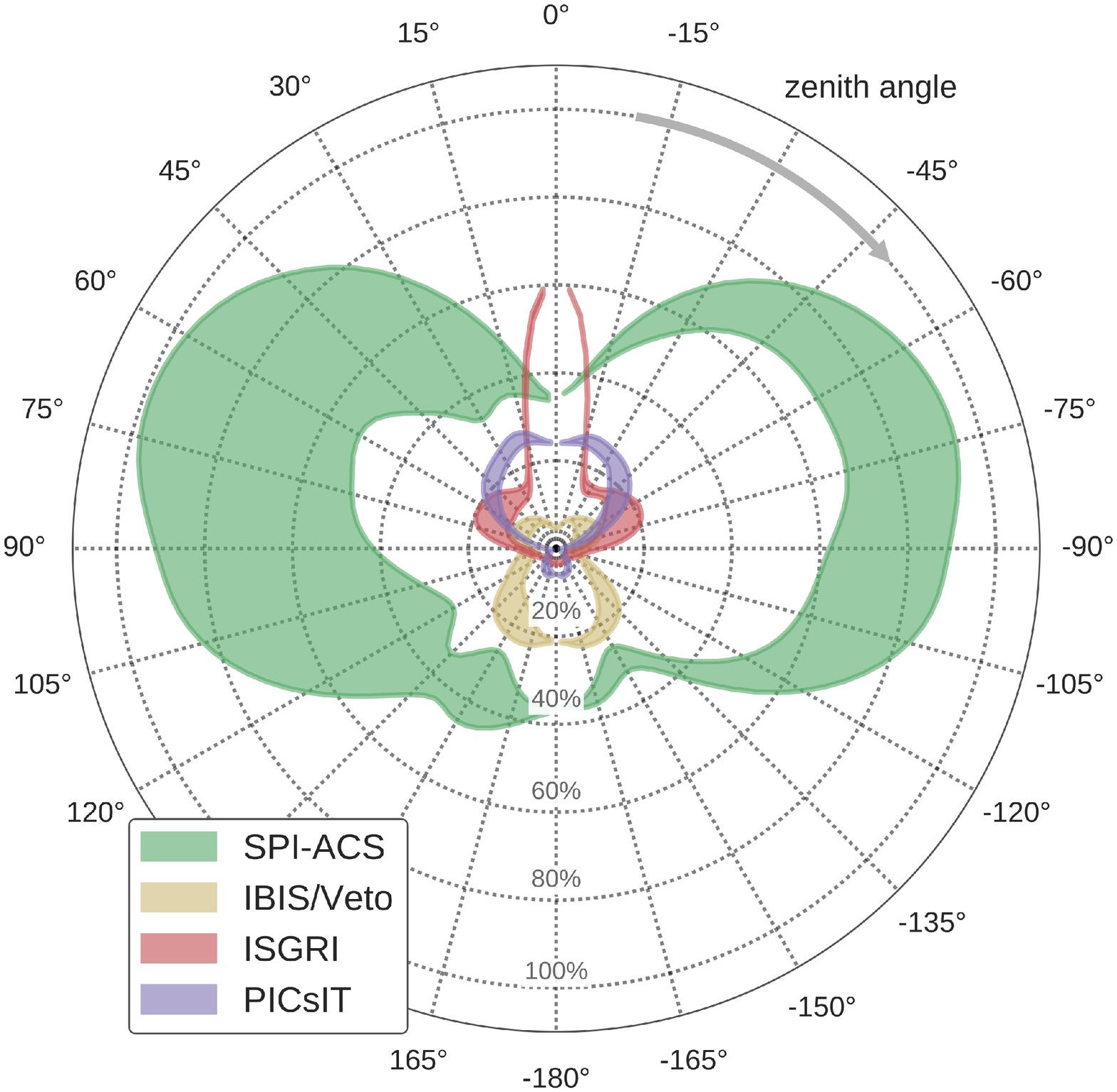}
  \includegraphics[width=\columnwidth]{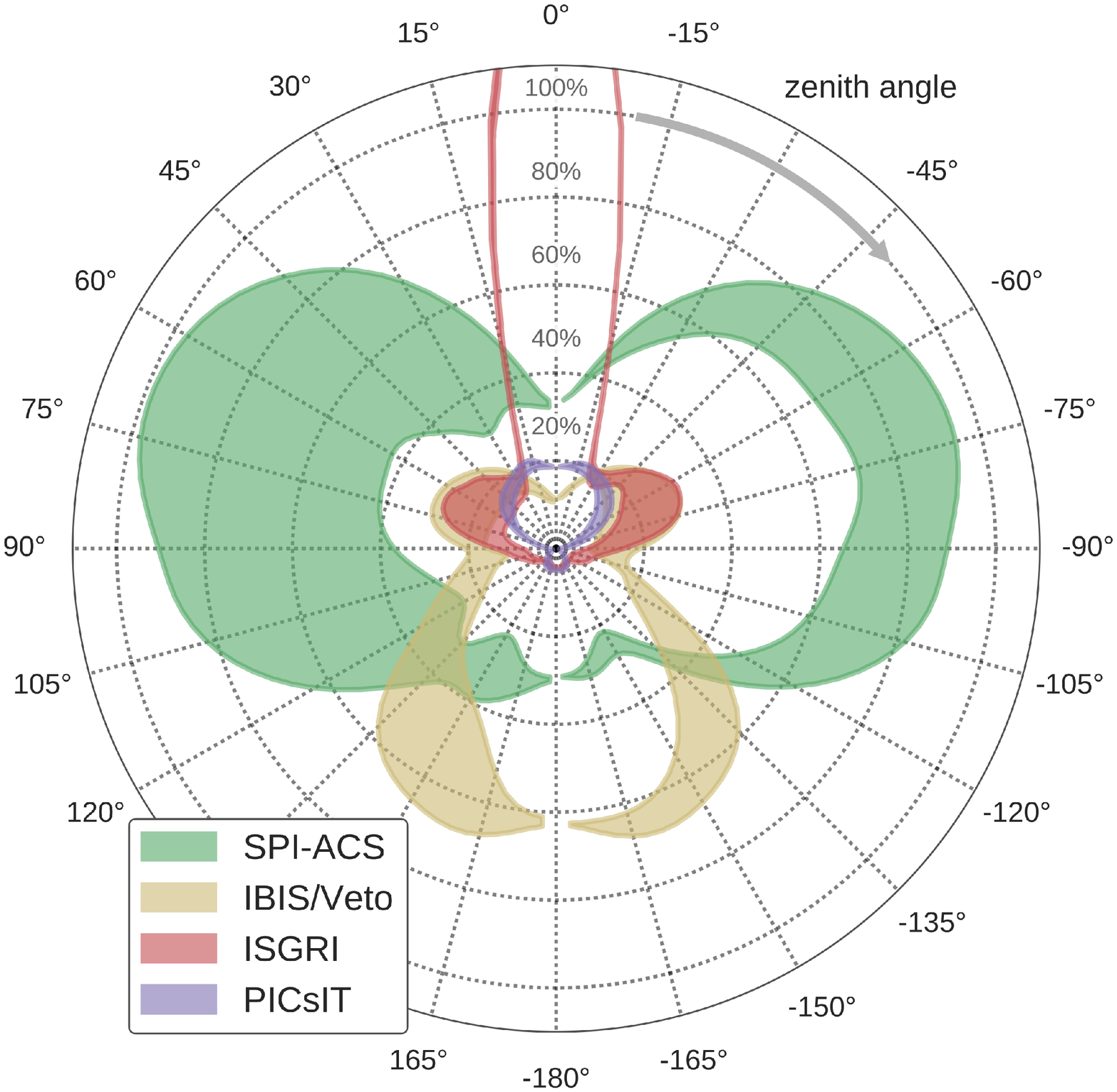}
  \includegraphics[width=1.5\columnwidth]{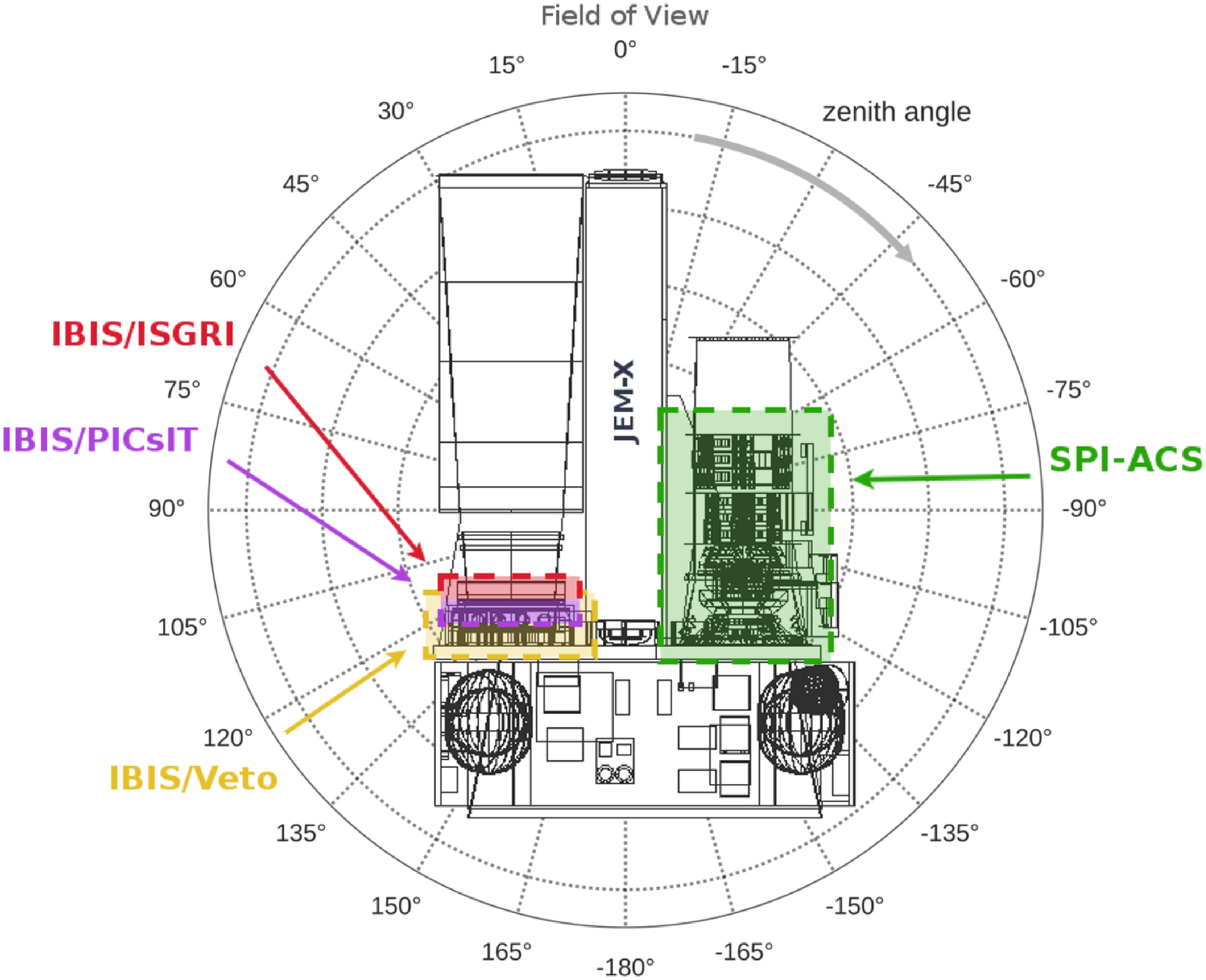}
  \caption{{\it Top left panel}: Plot of the detection significances achievable in the entire sky by the omni-directional instruments 
  on-board INTEGRAL, together with those of the IBIS/ISGRI and IBIS/PICsIT (see Figure~\ref{fig:sens_4fov} for a more detailed description of the 
   imaging instruments sensitivity for observations within their FoVs). The detection significances have been normalized to the maximum 
   value that can be obtained by the SPI-ACS in the most favorable orientation and considering  
   a 1~s long event which emission is characterized by a Comptonized 
   spectral model with $\alpha$=-0.5 and E$_p$=600~keV. 
  A shaded region is drawn for each instrument to represent how the detection significances change at different zenith angles  
  (explicitly marked on the figure axes), taking into account also the dependence of the instrument sensitivity as a function 
  of the azimuthal angles (not explicitly indicated in the figure). For each zenith angle, we computed the distribution of the 
  significances for all relevant azimuthal angles and excluded for clarity from the shaping of the shaded regions the 10\% worst 
 and 10\% best values.  The satellite pointing direction is toward zenith angle 0~deg, 
  while the axis pointing from SPI to IBIS is along the 90 deg direction. {\it Top right panel}: same as top left  
  but in the case of a 8~s-long event which emission is characterized by a Band spectral model \citep{band93} 
  with $\alpha$=-1, E$_p$=300~keV, and
  $\beta$=-2.5. {\it Bottom panel}: Sketch of INTEGRAL   
  derived from the INTEGRAL mass model (TIMM) showing the satellite pointing direction (zenith angle 0~deg), the axis connecting SPI to IBIS (zenith angle 
  90~deg), and the location of the different instruments.}
  \label{fig:isens_zenith_hard}
\end{figure*}

We used Monte-Carlo simulations exploiting the INTEGRAL mass model 
\citep{Ferguson2003} to compute the response of IBIS/Veto (see
Figure~\ref{fig:isens_zenith_hard}). 
We checked our results by using the observations of bright GRBs that were 
detected by Fermi/GBM. For a good match, we had to adjust  
the low energy threshold of the IBIS/Veto system for which we have a limited description. 
The estimated discrepancy between the observed number of counts in the IBIS/Veto and those expected 
based on the GBM results is found in all cases to be $\lesssim$20\%. 

IBIS/Veto is a particularly useful instrument to study sources with off-axis angles
larger than 120~deg, where the sensitivity of  ISGRI, PICsIT, and the IBIS Compton mode
are low. At these angles, also the coverage provided by the SPI-ACS is limited (see Figure~\ref{fig:sky_sens_4}). 
We also note that there is a relatively small fraction of the sky (about 15\%, depending on the source spectral 
energy distribution) for which the effective area of the IBIS/Veto is larger than that of the SPI-ACS. 
Owing to its lower background, the IBIS/Veto has a factor of 4 improved sensitivity  
compared to the SPI-ACS for the detection of impulsive events that are longer than 8~s and localized 
in some directions close to the opposite of satellite pointing axis (see Fig~\ref{fig:isens_zenith_hard}).

\subsection{SPI} \label{sec:spi}

SPI is a $\gamma$-ray spectrometer made of 19 hexagonal high-purity
Germanium detectors.  A tungsten coded aperture mask characterized by a hexagonal pattern
is located 1.7 m above the detection plane of the instrument, thus allowing SPI to image a large
region of the sky at once (the fully-coded FoV radius is $\sim$8~deg) with a typical angular 
resolution of 2.5~deg. 

The Germanium detectors are surrounded by an active anti-coincidence
shield, SPI-ACS, forming a collimator tube and a bottom shield. 
SPI-ACS is made of 91 BGO (Bismuth Germanate, Bi$_4$Ge$_3$O$_{12}$)
scintillator crystals and is very efficient in preventing 
$\gamma$-rays from reaching the germanium detectors from every direction
apart from the instrument FoV (see Sect.~\ref{sec:spiacs}). SPI data can thus only be exploited to
study sources within its FoV.

\subsubsection{SPI-ACS} \label{sec:spiacs}

Besides its main function of providing a veto signal for charged 
particles irradiating the SPI instrument, ACS is also able to 
provide the count-rate corresponding to all impinging particles and high energy photons. 
It can thus be used as a nearly omnidirectional detector of transient 
events with an effective area reaching 0.7~m$^2$ at energies above $\sim$75~keV 
and a time resolution of 50~ms \citep{spiacs}. 
The effective area of SPI-ACS is strongly affected by the opacity of 
materials, which are used for the INTEGRAL satellite structure and other instrument 
detectors. To estimate this
opacity, we performed a number of Monte-Carlo particle transport simulations using
both the SPI detector mass model \citep{sturner03} and the INTEGRAL satellite mass
model \citep[TIMM;][]{Ferguson2003}. We verified that both models  
predicted similar count-rates for the known impulsive events observed by SPI-ACS 
to within an accuracy of 20\%. For the detector response evaluation, we made 
use only of the TIMM, as it known to include a more accurate description of  spacecraft structures.

\bibliographystyle{aa}
\bibliography{references_merged}

\end{document}